\documentclass[sigconf, nonacm]{acmart}
\usepackage{pifont}
\usepackage{comment}
\usepackage{tikz}
\tikzset{every tree node/.style={align=center, anchor=north}}
\usetikzlibrary{trees,arrows.meta,chains,fit,shapes,calc}
\usepackage{balance}
\usepackage{subcaption}
\usepackage{tabularx}
\usepackage[export]{adjustbox}
\usepackage{diagbox}
\usepackage{hyperref}
\usepackage[noabbrev, capitalise, nameinlink]{cleveref}
\usepackage{xcolor,colortbl}
\usepackage[vlined,noend,linesnumbered,ruled,resetcount]{algorithm2e}
\usepackage{MnSymbol}
\usepackage{amsmath}
\usepackage{mathtools}
\usepackage{multirow,hhline}
\usepackage{enumitem}
\usepackage{multicol}
\let\oldnl\nl
\newcommand{\nonl}{\renewcommand{\nl}{\let\nl\oldnl}}
\definecolor{light-gray}{gray}{0.75}
\definecolor{light-light-gray}{gray}{0.8}
\definecolor{mid-green}{HTML}{E0F3DB}
\definecolor{mid-red}{HTML}{FBB4AE}

\newtheorem{definition}{Definition}

\newtheorem{lemma}{Lemma}

\newtheorem{theorem}{Theorem}
\newtheorem{myexp}{Example}

\newtheorem{proposition}{Proposition}

\def\eqref#1{equation~\ref{#1}}

\def\1{\bm{1}}

\newcommand{\schema}{\ensuremath{\mathcal{S}}}

\newcommand{\dom}{\ensuremath{Dom}}
\newcommand{\paratitle}[1]{\vspace{2mm} \noindent{\bf #1.}}
\newcommand{\admissible}{\ensuremath{A}}
\newcommand{\prot}{\ensuremath{P}}
\newcommand{\outcome}{\ensuremath{O}}
\newcommand{\prob}{FDPSynth}
\newcommand{\mst}{Marginals-MST}
\newcommand{\sys}{{\tt PreFair}} 

\DeclareMathAlphabet{\mathsfit}{\encodingdefault}{\sfdefault}{m}{sl}
\SetMathAlphabet{\mathsfit}{bold}{\encodingdefault}{\sfdefault}{bx}{n}

\def\gA{{\mathcal{A}}}

\def\gC{{\mathcal{C}}}

\def\gG{{\mathcal{G}}}

\def\gM{{\mathcal{M}}}
\def\gN{{\mathcal{N}}}

\def\gQ{{\mathcal{Q}}}

\SetCommentSty{mycommfont}

\newcommand{\reva}[1]{#1}
\newcommand{\revb}[1]{#1}
\newcommand{\revc}[1]{#1}
\newcommand{\common}[1]{#1}

\newcolumntype{Y}{>{\centering\arraybackslash}X}
\newlength{\Oldarrayrulewidth}

  \newcolumntype{s}{>{\hsize=.01\hsize}X}

\newcommand\vldbpagestyle{plain} 

\begin{document}
\title{PreFair: Privately Generating Justifiably Fair Synthetic Data}

\author{David Pujol}
\affiliation{%
  \institution{Duke University}
}
\email{dpujol@cs.duke.edu}

\author{Amir Gilad}
\affiliation{%
  \institution{Duke University}
}
\email{agilad@cs.duke.edu}

\author{Ashwin Machanavajjhala}
\affiliation{%
  \institution{Duke University}
}
\email{ashwin@cs.duke.edu}

\begin{abstract}

When a database is protected by Differential Privacy (DP), its usability is limited in scope. 
In this scenario, generating a synthetic version of the data that mimics the properties of the private data allows users to perform any operation on the synthetic data, while maintaining the privacy of the original data. 
Therefore, multiple works have been devoted to devising systems for DP synthetic data generation. 
However, such systems may preserve or even magnify properties of the data that make it unfair, rendering the synthetic data unfit for use. 
In this work, we present \sys, a system that allows for DP fair synthetic data generation. \sys\ extends the state-of-the-art DP data generation mechanisms by incorporating a causal fairness criterion that ensures fair synthetic data. 
We adapt the notion of justifiable fairness to fit the synthetic data generation scenario. We further study the problem of \revb{generating DP fair synthetic data}, showing its intractability and designing algorithms that are optimal under certain assumptions. 
We also provide an extensive experimental evaluation, showing that \sys\ generates synthetic data that is significantly fairer than the data generated by leading DP data generation mechanisms, while remaining faithful to the private data.
\end{abstract}

\maketitle
\setcounter{page}{1}
\pagestyle{\vldbpagestyle}



\section{Introduction}\label{Sec:Introduction}
Among the various privacy definitions that have been explored, Differential Privacy (DP) \cite{DiffPriv} has emerged as the most reliable. 
DP allows users to get answers to queries and even train Machine Learning models over private data by augmenting the true answers with noise that cloaks information about individuals in the data. Indeed, multiple works have focused on answering queries over private data \cite{Matrix,MWEM,PMW,HDMM, PrivateSQL} and training models over such data \cite{AbadiCGMMT016,PapernotAEGT17}. 
However, performing queries or training models directly on the private data has several prominent drawbacks. First, the class of allowed operations is limited. For example, only aggregate queries and specific techniques of training are allowed. Second, the data itself cannot be shared or used to explain the results and verify them without spending more precious privacy budget.

\begin{figure}[t]
\vspace{-3mm}
\begin{center}
\includegraphics[width = \linewidth]{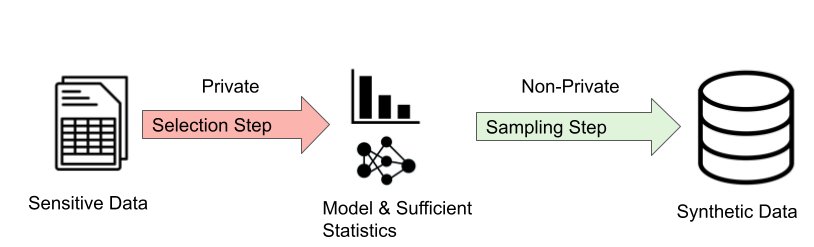}
\end{center}
    \caption{Popular private synthetic data framework design.}
    \label{fig:syth_framework}
\end{figure}

These limitations and others have driven the approach of generating synthetic data under DP \cite{ChenOF20,LiuV0UW21,Li21,LiXZJ14,AydoreBKKM0S21,GeMHI21,JordonYS19a,MWEM,LH21,SnokeS18,Torkzadehmahani19,Rosenblatt20,xie18,McKennaSM19,RyanMnist}. 
In general, these systems often work in two main steps as depicted in \cref{fig:syth_framework}. First, they measure some set of summary statistics and generate some model to represent the data that is consistent with the measured statistics. This step is done in a privacy preserving manner by infusing noise into both the statistics and the model generating step. From that point, the synthetic data is sampled directly from the model in a non-private manner. 
DP synthetic data provides a strong formal guarantee of privacy on the original private data while creating a synthetic dataset that retains many of the properties of the original dataset. The synthetic data may then be used in any way and shared without additional privacy risk, due to the post-processing property of DP \cite{dwork2014:textbook}. 

However, as indicated by \cite{GanevOC22}, the synthetic data generation process may preserve or even exacerbate properties in the data that make it unfair. 
There has been a significant effort in recent years to both define fairness \cite{proxyfair,InterFair,Counterfactual,VermaR18,Chouldechova17,Corbett-DaviesP17,DworkHPRZ12,Berk21,GalhotraBM17,KleinbergMR17,NabiS18}, and develop frameworks to understand and mitigate bias \cite{SelbstBFVV19,SureshG21,FeldmanFMSV15}. 
Among these, causal approaches for fairness \cite{InterFair,Counterfactual,proxyfair} propose a principled approach of measuring the effect of protected attributes on the outcome attributes. 
These works consider the causal relationship between protected attributes, i.e., those that cannot be explicitly used in decision making, admissible attributes, i.e., those that can be explicitly used in decision making, and outcome attributes that are the prediction targets. 

Prior works have proposed frameworks for pre-processing steps to repair the data and ensure causal fairness \cite{FeldmanFMSV15,CalmonWVRV17,InterFair}. 
Notably, none of these have done so while under the constraint of DP as they assume the data is accessible.

{\em In this paper, we propose \sys\ ({\bf Pr}ivat{\bf e} and {\bf Fair} synthetic data generation), a novel framework for generating fair synthetic data that satisfies DP.} 
Our solution combines the state-of-the-art DP synthetic data generation system, MST~\cite{RyanMnist} (shown to be the best performing mechanism in terms of minimizing the discrepancy between the private database and the generated one \cite{SynthBenchmark}) %
with the robust causal fairness definition of Justifiable Fairness \cite{InterFair} to generate data that ensures fairness while satisfying DP. In justifiable fairness, the data is defined to be fair if there is no directed path in the causal graph between the protected attribute and the outcome attribute that does not pass through an admissible attribute in a graphical model which describes the data. 
Specifically, \sys\ augments the first step shown in \cref{fig:syth_framework} to satisfy the graph properties needed for Justifiable Fairness and provides a guarantee that the data generated in the second step will be fair, under a common assumption.

\begin{myexp}\label{ex:intro}
Consider the Adult dataset \cite{adult} with the attributes age, marital-status (marital), education (edu), sex, income, and relationship. 
The MST system first learns the 2-way attribute marginals and then samples from them to generate a synthetic database instance.
 
\cref{fig:non_fair_MST} shows the graphical representation of the database attributes, generated by MST. Here, protected attributes are denoted with a subscript $\Pi$ (and are colored red), admissible attributes are denoted with a subscript $\alpha$ (and are colored blue) and the outcome attribute with a subscript $\Omega$ (and is colored green). 
The edges denote attributes that have a strong correlation between them in the private database. 
Note that the protected attribute sex (denoted by a subscript $\Pi$ and colored red) has an edge to the income attribute which is the outcome. 
Intuitively, any synthetic database that will follow the distribution defined by this graph will have a direct dependence between the attribute sex and the attribute income, \reva{resulting in unfair data}  
(while the graphical model is not a causal graph, we rely on the graphical properties of justifiable fairness). 
On the other hand, \cref{fig:fair_MST} shows an alternative graphical representation  of the dependencies, generated by \sys. In this graph, the path from sex to income passes through the admissible attribute education (the dashed edges colored blue). 
Intuitively, in any synthetic data that will adhere to this distribution, the influence of the sex attribute on income will be mitigated by the education attribute, which is considered admissible, so, it is allowed to directly influence the outcome. 
\end{myexp}

Ensuring that \sys\ is useful gives rise to several challenges. In particular, we tackle the following challenges: (1) adapting the definition of fairness to undirected graphical models, (2) making a connection between the graphical model and the data to guarantee that a fair model would lead to the generation of fair data, and (3) providing efficient algorithms that will generate fair data while remaining faithful to the private data and satisfying DP.

\begin{figure}[t]
\begin{subfigure}{.48\linewidth}
\begin{center}
\begin{tikzpicture}[scale=0.6,snode/.style = {shape=rectangle, rounded corners, draw, align=center, top color=white, bottom color=blue!20},
outcome/.style={shape=rectangle, rounded corners,thick,draw,, top color=white, bottom color=green!25},
none/.style={shape=rectangle, rounded corners,thick,draw,, top color=white, bottom color=light-gray}]

\begin{scope}[every node/.style={shape=rectangle, rounded corners,thick,draw,, top color=white, bottom color=blue!25}, 
other/.style={shape=rectangle, rounded corners,thick,draw, top color=white, bottom color=red!25}]
    \node (Q) at (2,5.5) {\footnotesize $marital_\alpha$};
    \node[none] (A) at (0,3) {\footnotesize $relation$};
    \node[none] (B) at (0,5.5) {\footnotesize $age$};
    \node[other] (F) at (2,3) {\footnotesize $sex_\Pi$};
    \node (P) at (4.5,3) {\footnotesize $edu_\alpha$};
    \node[outcome] (Q2) at (4.5,5.5) {\footnotesize $income_\Omega$};
\end{scope}

\begin{scope}[>={Stealth},
            standard/.style={draw=black, very thick}, 
              unfair/.style={draw=red, very thick, densely dashdotted}]
    \path [-,standard] (B) edge node {} (A);
    \path [-,standard] (Q) edge node {} (F);
    \path [-,standard] (B) edge node {} (Q);
    \path [-,standard] (P) edge node {} (F);
    \path [-,unfair] (Q2) edge node {} (F);
\end{scope}

\end{tikzpicture}
\end{center}
    \caption{Non-fair model gen. by \cite{RyanMnist}.}
    \label{fig:non_fair_MST}
\end{subfigure}%
\begin{subfigure}{.48\linewidth}
\begin{center}
\begin{tikzpicture}[scale=0.6]

\begin{scope}[every node/.style={shape=rectangle, rounded corners,thick,draw,, top color=white, bottom color=blue!25}, 
other/.style={shape=rectangle, rounded corners,thick,draw,, top color=white, bottom color=red!25},
outcome/.style={shape=rectangle, rounded corners,thick,draw,, top color=white, bottom color=green!25},
none/.style={shape=rectangle, rounded corners,thick,draw,, top color=white, bottom color=light-gray}
]
    \node (Q) at (2,5.5) {\footnotesize $marital_\alpha$};
    \node[none] (A) at (0,3) {\footnotesize $relation$};
    \node[none] (B) at (0,5.5) {\footnotesize $age$};
    \node (F) at (2,3) {\footnotesize $edu_\alpha$};
    \node[other] (P) at (4.5,3) {\footnotesize $sex_\Pi$};
    \node[outcome] (Q2) at (4.5,5.5) {\footnotesize $income_\Omega$};
\end{scope}

\begin{scope}[>={Stealth},
            standard/.style={draw=black, very thick}, 
              unfair/.style={draw=red, very thick, densely dashdotted},
              admissible/.style={draw=blue, very thick, dashed}]
    \path [-,standard] (B) edge node {} (A);
    \path [-,standard] (Q) edge node {} (F);
    \path [-,standard] (B) edge node {} (Q);
    \path [-,admissible] (P) edge node {} (F);
    \path [-,admissible] (Q2) edge node {} (F);
\end{scope}

\end{tikzpicture}
\end{center}
    \caption{Fair model gen. by \sys.}
    \label{fig:fair_MST}
\end{subfigure}
\caption{Graphical data generation models.}\label{fig:msts}
\end{figure}

\paratitle{Our contributions}
\begin{itemize}[topsep=0pt,itemsep=-1ex,partopsep=1ex,parsep=1ex, leftmargin=*]
    \item We propose \sys, the first framework, to our knowledge, that generates synthetic data that is guaranteed to be fair and differentially private. \sys\ combines the state-of-the-art synthetic data generation approach that satisfies DP \cite{RyanMnist} with the definition of justifiable fairness \cite{InterFair}. 
    \item We provide a novel model for fair data generation that relies on probabilistic graphical models and characterize the desiderata for the sampling approach to generate justifiably fair data. 
    \item  We prove that finding the optimal fair graph in our model is NP-hard even in the absence of privacy.  
    \item Due to the intractability of the problem, we devise two solutions: (1) an algorithm that provides an optimal solution for an asymptotically large privacy budget and (2) a greedy algorithm that uses heuristics to construct the graph, while still guaranteeing fairness but possibly reducing the faithfulness to the private data. 
    \item We perform an extensive experimental evaluation to test \sys\ and show that our greedy approach can generate fair data while preserving DP and remaining faithful to the private data. We further show that our greedy approach does not incur major overhead in terms of performance. 
\end{itemize}
\section{Preliminaries}\label{Sec:background}

\paratitle{Data}
We assume a single table schema $\schema = R(A_1, A_2 \dots A_d)$ where $\gA = \{A_1, A_2 \dots A_d\}$ denotes the set of attributes $R$. Each attribute $\gA_i$ has a finite domain $\dom(A_i)$. The full domain of $R$ is $\dom(R) = \dom(A_1) \times \dom(A_1) \times \dots \dom(A_d)$. 
An instance $D$ of relation $R$ is a bag whose elements are tuples in $\dom(R)$, i.e., a tuple can be written as $t = (a_1, \ldots, a_d)$ where $a_i \in \dom(A_i)$. 
The number of tuples in $D$ is denoted as $|D| = n$. 
We consider synthetic databases. Given a database $D$ over schema $\schema$ and $n$ tuples, we say that $D'$ is a synthetic copy of $D$ if $D'$ also has the schema \schema\ and contains $n$ tuples of the form $t = (a_1', \ldots, a_d')$ where $a_i' \in \dom(A_i)$.

For the rest of the paper, we refer to the process of generating a synthetic copy of a database $D$ as synthetic data generation. 

\subsection{Renyi Differential Privacy}
\label{sec:Renyi}
In this paper, we employ the notion of Renyi-DP (RDP) \cite{RenyiDP}. This is a formal model of privacy that guarantees each individual that any query computed from sensitive data would have been almost as likely as if the individual had opted out. More formally, RDP is a property of a randomized algorithm that bounds the Renyi Divergence of output distributions induced by changes in a single record. To define privacy, we begin with neighboring databases.

\begin{definition}[Neighboring Databases]
Two databases $D$ and $D'$ are considered neighboring databases if $D$ and $D'$ differ in at most one row. We denote this relationship by $D' \approx D$. 
\end{definition}

We often use the notion of neighboring databases to distinguish the impact of any particular individual's input on the output of a function. We likewise measure the maximum change in any function due to the removal or addition of one row often calling it the sensitivity of the function.

\begin{definition}[Sensitivity]\label{def:sensitivity}
Given a function $f$ the sensitivity of $f$ is $sup_{D' \approx D}{|f(D)-f(D')|}$ and is denoted by $\Delta f$.

\end{definition}

From here we can define our formal notion of Privacy, RDP.
\begin{definition}[Renyi Differential Privacy]\label{def:RDP}
A rand\-om\-ized mechanism $\gM$ satisfies $(\alpha, \gamma)$-RDP for $\alpha \geq 1$ and $\gamma \geq 0$ if for any two neighboring databases $D$, and $D'$:
\begin{equation*}
D_\alpha(\gM(D) \| \gM(D')) \leq \gamma 
\end{equation*}
where $D_\alpha ( \cdot \| \cdot )$ is the Renyi divergence of order $\alpha$ between two probability distributions.
\end{definition}
This ensures that no single individual contributes too much to the final output of the randomized algorithm by bounding the difference of the distributions when computed on neighboring databases.\par 
All mechanisms that satisfy RDP also satisfy classic differential privacy \cite{DiffPriv}. Given privacy parameters $\alpha$ and $\gamma$ the RDP guarantee can be translated into a classical DP guarantee as follows.
\begin{theorem}[RDP to DP \cite{RenyiDP}]\label{thrm:RDPtoDP}
 If a mechanism $\gM$ satisfies $(\alpha, \gamma)$-RDP it also satisfies $\left(\gamma + \frac{\log(1/\delta)}{\alpha -1}, \delta\right)$ - DP for all $\delta \in (0,1]$.
\end{theorem}

The parameter $\gamma$ quantifies the privacy loss. \revb{While classic DP bounds the worst-case privacy loss, RDP treats privacy loss as a random variable which allows for computing a tighter bound over privacy loss in many situations. Our mechanisms, such as Algorithm \ref{algo:fair_mst_2} (\cref{sec:optimal}), use repeated calls to the Gaussian mechanism and benefit from this tighter analysis.}
If there are two RDP releases of the same data with two different privacy parameters the amount of privacy loss is equivalent to the sum of their privacy parameters. 
\begin{theorem}[RDP composition \cite{RenyiDP}]
\label{thrm:RDPcomposoition}
Let $\gM_1$ be an $(\alpha, \gamma _1)$-RDP algorithm and $\gM_2$ be an $(\alpha, \gamma _1)$-differentially private algorithm. Then their combination defined to be $\gM_{1,2}(x) = (\gM_1(x), \gM_2(x))$ is $(\alpha, \gamma _1 + \gamma _2)$-RDP.
\end{theorem}

When designing private systems, designers are required to balance utility and privacy loss over several mechanisms to ensure that a global privacy loss is not exceeded. 
RDP is robust to post-processing, i.e., if any computation is done on an  RDP data release without access to the original data, the additional computation also satisfies RDP.

\paratitle{The Gaussian mechanism \cite{RenyiDP}} This is a DP primitive for answering numerical queries and is often used to construct more complex DP mechanisms.
\begin{definition}[Gaussian Mechanism]
Given a query  $q$ and a database, $D$ the randomized algorithm which outputs the following query answer is $(\alpha, \alpha \frac{\Delta q^2}{2\sigma^2})$-RDP \cite{RenyiDP} for all $\alpha \geq 1$. 
\begin{equation*}
q(D) + \gN(0,\sigma^2)
\end{equation*}
\end{definition}
where $\gN(0,\sigma)$ denotes a sample from the Normal distribution with mean $0$ and scale $\sigma$, and $\Delta q$ is the sensitivity of $q$ (\cref{def:sensitivity}).

\paratitle{The Exponential Mechanism (EM) \cite{ExpMech}} EM is an RDP primitive for queries that outputs categorical data instead of numerical data. The exponential mechanism releases a DP version of a categorical query $Q$ by randomly sampling from its output domain $\Omega$. The probability for each individual item to be sampled is given by a pre-specified score function $f$. The score function takes in as input a dataset $D$ and an element in the domain $\omega \in \Omega$ and outputs a numeric value that measures the quality of $\omega$. Larger values indicate that $\omega$ is a better output with respect to the database and as such increases the probability of $\omega$ being selected. More specifically given a dataset $D$ the exponential mechanism samples $\omega \in \Omega$ with probability proportional to $\exp(\frac{\epsilon}{2\Delta f} \cdot f(D,\omega))$ where $\Delta f$ is the sensitivity of the scoring function $f$.

\begin{theorem}[From \cite{ExpMech}]
\label{thrm:RDPexpmech}
The Exponential Mechanism applied to the quality score function $f$ satisfies $(\alpha, \alpha \frac{(2 \epsilon \Delta f)^2}{8})$- RDP. Where $\Delta f$ is the sensitivity of $f$
\end{theorem}

Both the Gaussian and Exponential mechanism are DP primitives which are often used as subroutines in the design of more complex DP mechanisms.

\subsection{\mst}\label{sec:prelim-mst}
Previous work \cite{RyanMnist} has proposed an approach of using a \mst\ for generating DP synthetic data.
The \mst\ is based on a Bayes network that encodes the 2-way marginals between the different attributes of the database. 
This approach was shown to be the state-of-the-art DP mechanism for generating synthetic data \cite{SynthBenchmark}. 
In the selection step (recall \cref{fig:syth_framework}), the system generates a \mst\ which maximizes the pairwise mutual information over the attributes of the database. 
The mutual information between two attributes measures the mutual dependence between the two attributes. Two attributes that depend on each other more have higher mutual information. As such the 2-way marginals of attributes with high dependence on one another are chosen to be measured directly. Mutual information is defined as follows.

\begin{definition}[Mutual Information \cite{cover_thomas_2005}]\label{def:error}
The mutual information between two attributes $A_i$ and $A_j$ is defined as follows.
\begin{small}
\begin{equation}
\hspace{-2mm}
    \sum_{a_i \in \\ dom(A_i)}\quad \quad \sum_{\mathclap{a_j \in\\ dom(A_j)}} P[A_i= a_i, A_j = a_j] \log\left(\frac{P[A_i= a_i, A_j = a_j]}{P[A_i= a_i]P[A_j = a_j]}\right)
\end{equation}
\end{small}
\end{definition}

\paratitle{Selection step} The \mst\ selection step is a three step process each with its own split of the privacy parameter. First, it measures the 1-way marginals. It then selects a set of 2-way marginals to measure by creating a graph where the nodes represent the attributes and the edges represent 2-way marginals between attributes. The weights of each edge are set to (an estimation of) the mutual information between the two attributes. It then uses a private mechanism to generate \revb{a maximum spanning tree} (\mst) over this graph by using a private version of Kruskal's algorithm \cite{Kruskal} where the exponential mechanism is used to select edges. 

\begin{definition}[\mst]
Given a database $D$ over the attribute set $\gA$ and an undirected graph $G = (V,E,w)$, where $V = \gA$, $E$ are edges between attributes, and $w:V\times V\to \mathbb{R}^+$ returns the mutual information between every pair of attributes such that $(A_1,A_2) \not\in E$ iff $w(A_1,A_2) = 0$, a \mst\ is a spanning tree of $G$.
\end{definition}

Once the marginals are selected the final graphical model is constructed using Private PGM \cite{McKennaSM19} which (approximately) preserves the conditional independence properties of the \mst\ as well as its marginals. From here the model is sampled directly.\par

\paratitle{Sampling step}
\common{Once the graphical model is generated, the data is sampled in a way similar to sampling from a Bayes net.}
It \reva{samples attributes} one at a time with probabilities dependent on its already sampled neighbors. This imposes a direction in all the edges in the \mst\ resulting in a probability distribution that can be encoded in a directed Bayes net. The direction of the edges is directly dependent on the order in which the attributes are sampled with edges going from nodes which were sampled first toward nodes which were sampled later.

\begin{figure}[h]

\begin{center}
\begin{tikzpicture}[scale=0.6]

\begin{scope}[every node/.style={shape=rectangle, rounded corners,thick,draw,, top color=white, bottom color=blue!25}, 
other/.style={shape=rectangle, rounded corners,thick,draw,, top color=white, bottom color=red!25},
outcome/.style={shape=rectangle, rounded corners,thick,draw,, top color=white, bottom color=green!25},
none/.style={shape=rectangle, rounded corners,thick,draw,, top color=white, bottom color=light-gray}]
    \node (Q) at (2.7,5.5) {\footnotesize $marital_\alpha$};
    \node[none] (A) at (0,3.5) {\footnotesize $relation$};
    \node[none] (B) at (0,5.5) {\footnotesize $age$};
    \node (F) at (2.7,3.5) {\footnotesize $edu_\alpha$};
    \node[other] (P) at (6,3.5) {\footnotesize $sex_\Pi$};
    \node[outcome] (Q2) at (6,5.5) {\footnotesize $income_\Omega$};
\end{scope}

\begin{scope}[>={Stealth},
            standard/.style={draw=black, very thick}, 
             unfair/.style={draw=red, thick}]
    \path [-,standard,->] (B) edge node {} (A);
    \path [-,standard,->] (Q) edge node {} (F);
    \path [-,standard,->] (B) edge node {} (Q);
    \path [-,standard,<-] (P) edge node {} (F);
    \path [-,standard,<-] (Q2) edge node {} (F);
\end{scope}

\end{tikzpicture}
\end{center}
    \caption{Directed \mst\ as a result of sampling.}
    \label{fig:MST_directed}
\end{figure}
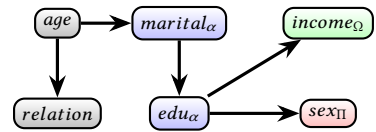

\begin{myexp}

\reva{\cref{fig:MST_directed} shows the process of sampling from the \mst\ shown in \cref{fig:fair_MST}}. 
The first attribute to be sampled is age. Its probability is dependent only on its 1-way marginal probability. Then, marital-status is sampled. Since age is already sampled, marital-status is then sampled according to the estimated 2-way marginal distribution. This process continues down the tree until all attributes are sampled.
\end{myexp}

Likewise, this ensures that the final distribution (approximately) follows the local Markov property of Bayesian networks.

\begin{definition}[Local Markov Property]
\label{def:local_Markov}
A graphical model $\gN$ satisfies the Local Markov Property if each attribute is conditionally independent of its non-descendants given its parent values. This is formulated as follows. 
\begin{equation} 
\forall a \in \gA: \space X_a\perp\!\!\!\perp X_{\gA \setminus \delta(a)} | X_{\Pi(a)}    
\end{equation}
\end{definition}

\revb{Where $\delta(a)$ are the descendants of $a$ and $\Pi(a)$ is the parents of $a$}. In particular, given the local Markov property, we can write the probability distribution on the attributes $\gA$ as follows. 
\begin{equation} \label{eq:bayes_prob}
    Pr[\gA] = Pr[X_1, X_2 \dots X_{|\gA|}]  = \prod_{i=1}^{|\gA|} = Pr[X_i|\Pi_i]
\end{equation}

where $\Pi_i$ is the parent set of $X_i$.
Given a Bayesian network a common inference task is determining conditional independence between two attributes. 

A sufficient criteria to denote conditional independence is d-separation \cite{causality} a condition that can be checked directly on the graph structure. 
We say that if $\gG$ is a graphical model and $Pr$ is a probability distribution then $Pr$ is Markov compatible with $\gG$ if a d-separation in $\gG$ implies conditional independence in $Pr$.
or more formally.
\begin{equation} \label{eq:Markov_compat}
   \mathbf{X} \perp\!\!\!\perp \mathbf{Y}|_d\mathbf{Z} \rightarrow \mathbf{X} \perp\!\!\!\perp \mathbf{Y}|\mathbf{Z} 
\end{equation}
If the converse is true, we say that $Pr$ is faithful to $\gG$. In terms of Bayesian networks, the following is true. 
\begin{theorem}
If $\gG$ is a directed acyclic graph over attributes $\gA$ and $Pr$ is a probability distribution that follows \cref{eq:bayes_prob}, then $Pr$ is Markov compatible with $\gG$.
\end{theorem}
 Interventions on a Bayseian network are achieved via the \textbf{do} operator denoted $\text{do}(X = x)$. On the graph structure itself this is equivalent to setting the value of a particular node and removing all the edges between that node and its parents. There is an equivalent notion on the distribution itself which can be computed analytically.

\subsection{Causal Fairness}
Causal fairness notions \cite{Counterfactual,proxyfair} consider the dependence between a set of protected attributes $P$ and a set of outcome decisions $O$. Protected attributes are those which are considered sensitive \revc{and} unactionable. For example, race and gender are considered unactionable attributes in highly sensitive tasks such as college admissions and loan applications. The outcome attribute $O$ is the objective to predict given the other attributes. 

Here, we focus on justifiable fairness \cite{InterFair} which considers \revb{additional admissible attributes}. These attributes are considered actionable for decision making despite their possible causal links to protected attributes. 

We begin by first introducing $K$ fairness and then using that to derive the notion of justifiable fairness.

\begin{definition}[K-fairness \cite{InterFair}]\label{def:k-fair}
Given the disjoint set of attributes $\gA$, protected attributes $\prot$, and outcome attribute $\outcome$ we say that a mechanism $\gM$ is $K$-fair with respect to 

$\prot$
if for any context $K= k$ and every outcome $O=o$ and every $P_i \in P$ the following holds: 
$$P[O=o|do(P_i=0), do(K=k)]=  P[O=o|do(P_i=1), do(K=k)]$$
\end{definition}

That is, conditioned on the intervention on the attributes in $K$, the value of the protected attribute ($P_i$) should not affect the final outcome ($O=o$).

K-fairness is usually defined with respect to the outcome of some algorithm such as a classifier but as most pre-processing techniques do, we will make the ``reasonable classifier'' assumption, where the conditional probability of classification is close to that same conditional probability of the training label in the training set. Therefore, we will use the outcome of a classifier and the training label interchangeably here.

\begin{definition}[Justifiable Fairness \cite{InterFair}]\label{def:justifiable-fair}
Let $\gA$ be the set \reva{of} all attributes, let $\prot \subset \gA$ be \revc{a set of} protected attributes, let $\admissible \subset \gA \setminus \prot$ be a set of admissible attributes, and let $O\in \gA \setminus (\prot \cup \admissible)$ be an outcome attribute\reva{.} A mechanism $\gM$ is justifiably fair if it is $K$-fair with respect to all supersets $K \supseteq A$.
\end{definition}

Prior work \cite{InterFair} notes that, for a directed causal graph $\gG$ over a probability distribution, if all directed paths from a protected attribute to the outcome attribute pass through at least one admissible attribute, then that probability distribution is justifiably fair.

\begin{theorem}[\cite{InterFair}]\label{thrm:bayes_path_original}
If all paths in a causal directed acyclic graph $\gG$ going from protected attributes to outcome attributes go through at least one admissible variable then $\gG$ is justifiably fair. If the probability distribution is faithful to the causal DAG, then the converse also holds.
\end{theorem}

While this is true for data that is faithful to a causal DAG, we will show in the sequel (\cref{prop:synth-fair}) that it is also true for data that is Markov compatible with that same graphical structure.

\section{A Framework for the Generation of Fair and DP Synthetic Data}\label{sec:model}
All of our mechanisms can be seen as imposing a distribution represented in the created Bayes net onto a particular dataset. This does not inherently preserve any causal structure nor impose any causal structure on the dataset it merely preserves a chosen set of measured statistics. In our setting, this is required as it is known that no private mechanism can preserve all possible statistics on a dataset \cite{dinur_nissim_2003} but instead, it must choose to preserve a subset of them and, even then, noise has to be infused into that subset. 

\subsection{Model For a Fair \mst}\label{sec:model-fair}
We now detail the model for a fair \mst, relate it to the synthetic data sampled in the second step \revc{(\cref{fig:syth_framework})}, and define the main problem at the center of our model.

\paratitle{Causal interpretation}
Previous work on justifiable fairness \cite{InterFair} satisfies justifiable fairness by changing the underlying distribution to make attributes \revb{that could be causally linked to instead be independent}. Fundamentally, this repair mechanism imposes independence between attributes and does not impose new casual structures. Similarly, our work can be seen as imposing an entire distribution onto a particular dataset. We generate distributions with independence properties that ensure the independence between the protected and outcome attributes when conditioned on the admissible attributes. We do not impose any given causal structures. Instead, we create data with a specific distribution from which no causality between protected and outcome attributes can be inferred.

\paratitle{Fairness of a \mst}
We can adapt the notion of interventional fairness to \mst. 
As a first step, we need to bridge the gap between an undirected \mst\ and \cref{def:k-fair} that assumes a DAG. 
Here, we consider the option of multiple outcomes rather than a single outcome attribute. This generalization is very practical for the synthetic data generation scenario, as we explain in the discussion in \cref{sec:problem}.

\begin{proposition}\label{thrm:bayes_path} 
Let $D$ be a database over the attribute set $\gA$, let $\prot,\admissible,\outcome \subseteq \gA$ be disjoint sets  representing the protected, admissible and outcome attributes respectively, and let $T$ be a directed \mst\ over $\gA$. If all directed paths in $T$ going from any attribute in $\prot$ to any outcome attribute in $\outcome$ go through at least one attribute in $\admissible$, then the distribution of $T$ is justifiably fair. 
\end{proposition}

This proposition leads to the definition of a fair \mst\ $T$ as one for which any directed \mst \ derived by directing the edges of $T$ satisfies the premise in \cref{thrm:bayes_path}.

\begin{definition}[Fair \mst]\label{def:fair-mst}
Given a database $D$ over attributes $\gA$, and disjoint sets $\prot,\admissible,\outcome \subseteq \gA$ representing the protected, admissible and outcome attributes respectively, a \mst, $T$, whose node set is $\gA$ is fair if in any directed \mst\ obtained by directing the edges of $T$, all directed paths in $T$ going from any attribute in $\prot$ to any outcome attribute in \outcome\ go through at least one attribute in $\admissible$.
\end{definition}

\begin{myexp}\label{ex:fair-mst}
\cref{fig:fair_MST} shows an example of a fair \mst. The only path from the protected attribute (sex) goes through at least one admissible attribute \reva{(education)}. Regardless of any direction imposed on the edges, the path between the outcome and protected attribute remains blocked by an admissible attribute.
\end{myexp}

\revb{We define a fair \mst\ as a \mst\ that satisfies \cref{thrm:bayes_path}, regardless of the imposed direction of edges. 
This ensures that a fair \mst\ represents a distribution that is justifiably fair 
regardless of the sampling order.} 
Since a \mst\ is a tree, we get the acyclic property `for free' as there cannot be cycles in the tree, regardless of the direction of the edges.

\paratitle{From a fair \mst\ to fair synthetic data}
We now show that, given a fair \mst, one can sample data which satisfies the property in \cref{def:justifiable-fair} as well, so long as the sampling approach preserves the distribution of the fair \mst. 

\begin{proposition}\label{prop:synth-fair}
Any synthetic database whose distribution is Markov Compatible (\cref{eq:Markov_compat}) with a directed fair \mst\ satisfies Justifiable Fairness (\cref{def:justifiable-fair}).
\end{proposition}

Intuitively, this proposition states that it is sufficient to find a fair \mst\ to ensure that the data derived from it is justifiably fair, as long as the sampling method is faithful to the distribution described by the fair tree. 
This is inherently true for any sampling method that preserves the distribution of a Bayes net (\cref{eq:bayes_prob}) such as the MST~\cite{RyanMnist} and PrivBayes~\cite{PrivBays} sampling steps.

\subsection{Problem Definition and Intractability}\label{sec:problem}

We have adapted the notion of interventional fairness to \mst\ and have further determined that it is sufficient to consider the fairness of the \mst\ to know that a synthetic database that is generated by a faithful sampling process will be fair.
We can now define \prob\ (Fair DP Synthetic Data Generation) as the problem at the heart of \sys. 

\begin{definition}[The \prob\ Problem]
Let $D$ be a database with attribute set $\gA$, a set of protected, admissible, and outcome attributes $\prot,\admissible,\outcome \subseteq \gA$, such that $\prot,\admissible,\outcome$ are \common{non-empty} and  pairwise disjoint. Let $G = (V,E,w)$ be a graph encoding the mutual information between every pair of attributes in $\gA$, where $V = \gA$ is the set of nodes, \common{$E = V \times V$} is the set of edges, and $w:V\times V \to \mathbb{R}^+ \common{\cup \{0\}}$ is a weight function for pairs of nodes that denotes the mutual information between every pair of attributes in $D$, $w(A_1,A_2) = 0$ iff $(A_1,A_2) \not\in E$ and the mutual information between $A_1$ and $A_2$ is $0$. 
The \prob\ problem is to find a fair MST that maximizes the pairwise mutual information between the attributes in $\gA$ while satisfying RDP \revb{for a given privacy parameter $\rho$}. 
\end{definition}

\begin{myexp}
Reconsider \cref{ex:intro}, where the protected attribute is sex, the admissible attributes are  education, marital-status and the outcome attribute is income. We assume that there is a graph where the edges represent the mutual information between these attributes in the private database. 
The \prob\ problem is to generate a fair \mst\ that maximizes the pairwise mutual information between the attributes while satisfying RDP for a given $\rho$. 
\end{myexp}

\common{We next give a simple existence claim, which will be implicitly used by our solutions (\cref{sec:algorithms}) to find a fair \mst. 
\begin{proposition}\label{prop:existance}

For any instance of the \prob\ problem, there exists a solution. 
\end{proposition}

}

\paratitle{The need for multiple outcomes}
Previous work on justifiable fairness \cite{InterFair} assumes that the database contains a single outcome attribute. However, in the context of generating private synthetic data this assumption is restrictive. In the non-private setting several different synthetic data sets can be constructed for individual tasks with different outcomes. In the private setting due to \cref{thrm:RDPcomposoition} each release of privacy protected data comes at the cost of additional privacy leakage. As a result, it is more efficient to release a single set of synthetic data which considers multiple outcomes.

\begin{myexp}
Reconsider \cref{fig:fair_MST} with the attribute relation as a second outcome. 
This simulates the case where the datatset needs to be used for two independent prediction tasks. 
In the first prediction task, income is to be predicted and in the second task relation status is to be predicted. Since both relation and income are to be predicted they should both be dependent on the admissible attributes and not the protected attribute (sex). 
\end{myexp}

\paratitle{NP-hardness of \prob}
Here, we show that \prob\ is intractable. We first define the decision version of the \prob\ problem without the RDP requirement. We show that, even in the absence of the RDP requirement, the decision problem is NP-hard. 
In the presence of private noise every instance of the problem must have non-zero probability, therefore, the addition of privacy still allows for the graph instances that are shown in the proof.

\begin{definition}[Decision version of \prob] \label{def:dec_prob}
Let $D$ be a database with attribute set $\gA$, a set of protected, admissible, and outcome attributes $\prot,\admissible,\outcome \subseteq \gA$, such that $\prot,\admissible,\outcome$ are \common{non-empty} and pairwise disjoint, and let $k \in \mathbb{N}$. Let $G = (V,E,w)$ be a graph encoding the mutual information between every pair of attributes in $\gA$, where $V = \gA$ is the set of nodes, \common{$E = V\times V$} is the set of edges, and $w:V\times V \to \mathbb{R}^+\common{\cup \{0\}}$ is a weight function for pairs of nodes that denotes the mutual information between every pair of attributes in $D$, $w(A_1,A_2) = 0$ iff $(A_1,A_2) \not\in E$ and the mutual information between $A_1$ and $A_2$ is $0$. 
The goal is to decide if there exists a fair \mst, $T = (V,E,w)$ of $G$ such that $\sum_{e\in E} w(e) \geq k$. 
\end{definition}

\common{
We now detail the NP-hardness claim and the reduction from 3-SAT leading to its proof. } 
\begin{proposition}\label{prop:nphard}
The decision version of \prob\ is NP-Hard.

\end{proposition}

\begin{proof}[Proof Sketch]
We show that, given an instance of 3-SAT,  we can define an equivalent instance of the \prob\ problem, where any maximal spanning tree will correspond to a satisfying assignment to all 3-CNF clauses. \reva{We assume, without loss of generality \cite{NMSAT}, that the 3-SAT instance has no trivial clauses and, for each literal, the formula contains at least one clause with that literal and at least one clause with its negation.}

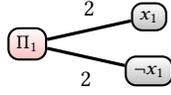
\begin{figure}
\begin{center}
\begin{tikzpicture}[scale=0.6]
\begin{scope}[every node/.style={shape=rectangle, rounded corners,thick,draw,, top color=white, bottom color=blue!25}, 
other/.style={shape=rectangle, rounded corners,thick,draw,, top color=white, bottom color=red!25},
none/.style={shape=rectangle, rounded corners,thick,draw,, top color=white, bottom color=light-gray}]
    \node[none] (X1) at (2.7,4.6) {\footnotesize $x_1$};
    \node[other] (P) at (0,4) {\footnotesize $\Pi_1$};
    \node[none] (X2) at (2.7,3.4) {\footnotesize $\neg x_1$};
\end{scope}

\begin{scope}[>={Stealth},
            standard/.style={draw=black, very thick}, 
             unfair/.style={draw=red, thick}]
    \path [-,standard,-] (P) edge node[above=2pt] {$2$} (X1);
    \path [-,standard,-] (P) edge node[below=2pt] {$2$} (X2);
\end{scope}
\end{tikzpicture}
\end{center}
    \caption{Assignment gadget.}
    \label{fig:assign}
    \vspace{-6mm}
\end{figure}

\begin{figure}
    \begin{center}
\begin{tikzpicture}[scale=0.6]
\begin{scope}[every node/.style={shape=rectangle, rounded corners,thick,draw,, top color=white, bottom color=blue!25}, 
other/.style={shape=rectangle, rounded corners,thick,draw,, top color=white, bottom color=red!25},
outcome/.style={shape=rectangle, rounded corners,thick,draw,, top color=white, bottom color=green!25},
none/.style={shape=rectangle, rounded corners,thick,draw,, top color=white, bottom color=light-gray}]
    \node[none] (X1) at (0,4.9) {\footnotesize $x_1$};
    \node (A) at (2.7,3.9) {\footnotesize $\alpha$};
    \node[none] (X2) at (0,3.1) {\footnotesize $x_2$};
    \node[none] (O) at (5,3.9) {\footnotesize $\outcome$};
    
    \node[none] (X') at (7,3.9) {\footnotesize $x'$};
    \node (A2) at (9.7,2.9) {\footnotesize $\alpha '$};
    \node[none] (X3) at (7,2) {\footnotesize $x_3$};
    \node[outcome] (O') at (12,2.9) {\footnotesize $\Omega_i$};
\end{scope}

\begin{scope}[>={Stealth},
            standard/.style={draw=black, very thick}, unfair/.style={draw=red, very thick, densely dashdotted},
              admissible/.style={draw=blue, very thick, dashed}]
    \path [-,admissible] (X1) edge node[above=0pt] {$2$} (A);
    \path [-,admissible] (X2) edge node[below=0pt] {$2$} (A);
    \path [-,admissible,-,bend left=20] (X1) edge node[above=2pt] {$1$} (O);
    \path [-,standard,-,bend right=20] (X2) edge node[below=2pt] {$1$} (O);
    
    \path [-,admissible,-] (X') edge node[above=0pt] {$2$} (A2);
    \path [-,admissible,-] (X3) edge node[below=0pt] {$2$} (A2);
    \path [-,admissible,-,bend left=20] (X') edge node[above=2pt] {$1$} (O');
    \path [-,standard,-,bend right=20] (X3) edge node[below=2pt] {$1$} (O');
    
    \path [-,admissible,-] (O) edge node[above=2pt] {$3$} (X');
\end{scope}

\end{tikzpicture}
\end{center}
    \caption{3-way OR gadget for a 3CNF clause $C_i$. The blue dashed edges represent the maximum spanning tree when there is an unblocked path from a protected attribute to $x_2,x_3$.}
    \label{fig:3_OR}
\end{figure}
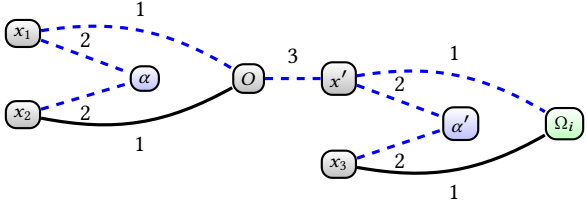

\paratitle{Graph construction}
We describe the construction of a graph $G$ from a 3-SAT instance $\varphi$ with $n$ literals and $m$ clauses.

\begin{enumerate}[leftmargin=*]
    \item  \reva{For each literal $x_i$ in $\varphi$, create an assignment gadget (see \cref{fig:assign}). This gadget contains a protected attribute $\Pi_i$ and edges from the protected attribute to two additional attributes representing the literal and its negation ($x_i$ and $\neg x_i$ respectively).}

    \item \reva{For each clause $C_i$, in $\varphi$ we create a 3-way OR gadget (see \cref{fig:3_OR}). The $\alpha$ and $\alpha'$ nodes in the OR gadget are admissible attributes and the $\Omega_i$ nodes are outcome attributes. The nodes $x_1, x_2, x_3$ are the inputs to the OR gadget and represent the literals in the corresponding clause.}
    
    \item \reva{Each of the inputs for the OR gate ($x_1,x_2,x_3$) are then connected to the corresponding literal in the assignment gadget using a weight 3 edge. Due to the constraints, both the literal and its negation will always connect to at least one OR gadget. E.g., the $x_1$ node in \cref{fig:3_OR} would be connected to the $x_1$ node in \cref{fig:assign} through a weight 3 edge\footnote{\reva{Edges that are not explicitly listed in the construction have weight 0 and can be ignored.}}.}
\end{enumerate}

We set \reva{$k = 22m + 2n$}, and thus $G = (V,E,w), k$ is an instance of the \prob\ problem. 
Given $\varphi$, this instance can be constructed in polynomial time and, therefore, the reduction is valid.

\paratitle{Proof of equivalence}
We now prove that $\varphi$ has a satisfying assignment iff there is a \mst\ with weight $\geq  22m + 2n$. 

\reva{$(\Rightarrow)$ 
First, we describe the construction of a fair \mst, $T$, given a satisfying assignment to $\varphi$, $\theta$. 

For each assignment gadget, add to $T$ only the edge corresponding to the literal set to $False$ by $\theta$. That is, $\theta(x_j) = False$ implies that $(\Pi_i, x_j)$ is included in $T$.  
For example, in \cref{fig:assign}, if $\theta(x_1) = True$, the edge $(\Pi_1, \neg x_1)$ would be added to $T$. 
All the weight 3 edge connecting the assignment gadgets and OR gadgets will always be added to $T$.
Then, for each OR gadget we add to $T$ edges such that any input which is set to $False$ has its path to $\Omega$ blocked by the admissible attributes. In \cref{fig:3_OR}, the blue dashed edges are added to $T$ when $\theta(x_1)=True$ and $\theta(x_2)=\theta(x_3)=False$. Note that any subtree of maximum weight ($13$) in the OR gadget can have at most two inputs whose path to $\Omega$ is blocked by an admissible attribute (these correspond to the literals set to $False$ by $\theta$). \par 

We will now show that $T$ is a fair \mst. By \cref{def:fair-mst}, we need to show that in any directed \mst, all directed paths in $T$ starting from $\Pi_i$ and ending at $\Omega_i$, go through an admissible node $\alpha$ or $\alpha'$. Since $\theta$ is a satisfying assignment, at most two of the inputs to each OR gadget have an unblocked path to a protected attribute (those set to $False$ by $\theta$). Since a maximum weight subtree of an OR gadget can block at most two paths from the inputs to $\Omega$, a set of edges can always be chosen such that all paths between protected and outcome attributes are blocked by admissible attributes. Since this holds for undirected paths it also holds for any direction given to those edges. For instance, the blue dashed edges in \cref{fig:3_OR} form a 
fair \mst\ when $x_1$ is not connected to a protected attribute.
This example is a valid setting in \textbf{any} assignment where $x_1$ is set to $True$ since $x_1$ has a direct path to $\Omega$ and the paths from $x_2, x_3$ are blocked by admissible attributes.

We now show that $T$ has a weight of $22m + 2n$.
This weight arises from choosing the tree of maximum weight ($13$) from each OR gadget (so far, the weight is $13m$) and a weight $3$ edge connecting the OR gadgets to the assignment gadget for each literal in the clause. Each clause has three such edges (so far, the weight is $13m + 3\cdot 3m$). Then, one edge is taken from each assignment gadget of weight $2$. Thus, the overall weight is $13m + 3\cdot 3m + 2n = 22m + 2n$. This is the maximum weight for any possible \mst\ since we chose all the maximum subtree for each of the gadgets.}

\reva{$(\Leftarrow)$
 
We now show how to convert a fair \mst, $T$, with weight $\geq  22m + 2n$ to an assignment, $\theta$, such that $\theta(\varphi) = True$.

To infer a satisfying assignment, $\theta$, from $T$, one only needs to consider the edges from the assignment gadgets. The chosen edge in each assignment gadget defines which literal is set to $False$, i.e., if $(x_i, \Pi_i) \in E(T)$ then $\theta(x_i) = False$. For example, in \cref{fig:assign}, if the edge 

$(x_1,\Pi_1)$ 
is chosen, then $\theta(\neg x_1) = False$ and $\theta(x_1)=True$. 

We now show that this assignment satisfies $\varphi$. For this, it is enough to show that any OR clause has at least one literal that is assigned to $True$ by $\theta$. 
Since $T$ is a fair \mst, the paths from the protected attributes  to the outcome attributes are blocked by admissible attributes in the OR gadgets. Each of the OR gadgets can block two of the paths from the inputs to the outcome attribute. Therefore, each OR gadget must have exactly one input that has an unblocked path to the outcome attribute. Since $T$ is fair, that attribute does not have unblocked paths to a protected attribute 
(this input is set to $True$ by $\theta$). 
Since this must be true for every OR gadget and each gadget corresponds to one clause, $\theta$ assigns at least one literal to $True$ in each clause and, hence, is a satisfying assignment for $\varphi$. 
In \cref{fig:3_OR}, since $T$ is fair, only $x_1$ has an unblocked path to $\Omega_i$, and this corresponds to $\theta(x_1) = True$. This same subtree is consistent with any assignment $\theta$ such that $\theta(x_1) =True$. $\theta(x_2)$ and $\theta(x_3)$ are not restricted by this gadget.}

\end{proof}

In order to connect the problem on an abstract graph to a database, we note that there always exists a database where the mutual information between attributes results in the structures of the assignment gadget and the 3-way OR gadget.

\begin{lemma}\label{lemma:database_existance}
There exists a database $D$ whose attributes have the mutual information relationship that results in the graph of \cref{fig:3_OR}.
\end{lemma}

\section{Computing a Fair \mst}\label{sec:algorithms}
Here we propose both an exponential time optimal solution as well as a linear time greedy algorithm. It is important to note when designing each of these mechanisms, we first designed a non-private solution to the fair \mst\ problem then adapt them to satisfy RDP. Even if an optimal non-private solution is adapted it does not ensure that the resulting private solution is also optimal. 
\subsection{\reva{Exponential} Asymptotically Optimal Algorithm}
\label{sec:optimal}

 \begin{algorithm}[h]
\SetAlgoLined

\caption{\reva{Exponential-\sys}}
\label{algo:fair_mst_2}
\SetKwInOut{Input}{input}\SetKwInOut{Output}{output}
\Input{Database D, set of admissible attributes A, set of protected attributes P, set of outcome attribute O, 
measurements of 1-way marginals $log$,
and a privacy parameter $\rho$}
\Output{Set of $(i,j)$ pairs to measure $\gC$}
Use Private-PGM\cite{McKennaSM19} to estimate all 2-way marginals $\bar{M}_{i,j}$ from $log$\\
Let $r$ be the number of 2 way marginals \\
Let $\sigma= \sqrt{\frac{ \rho}{2r}}$\\
Compute noisy measurements of $L_1$ error between the estimated 2-way marginal and real 2-way marginal for all $i,j$ pairs. $q_{i,j} = \|M_{i,j}(D) -\bar{M}_{i,j}\|$ using the Gaussian Mechanism with scale  $\sigma$ \label{line:exp_2_way}\\
Initialize empty priority queue $\gQ$ sorted in ascending order \label{line:queue}\\
Let $q_{max}$ be the maximum measurement. \label{line:max_measure}\\
Set the weight of each edge (i,j) to $\common{q_{max}} - q_{i,j}$ \label{line:weight} \\
Add the graph $ G = (\gA, \emptyset)$ to $\gQ$ \label{line:graph}\\
\While{$\gQ$ is non-empty \label{line:start_while}}{ 
Let $G = (\gA,\gC)$ be the first graph in $\gQ$\\
\If{$G$ is a \revb{spanning tree} \label{l:spanning}}{\Return $G$ \label{l:ret-tree}}
\revb{
\ForEach{edge $(i,j)$ that connects two connected components in $G$ and does not create an unblocked path from a node in $O$ to a node in $P$ \label{l:inner-for}}{add $G + (i,j)$ to $\gQ$ \label{l:add-edge}} 
}
\label{line:end_while}
}
\end{algorithm}

Here we present an exponential time solution for finding the optimal tree. The non-private version of this solution acts similarly to Dijkstra's algorithm. We take in as input the set of measured 1-way marginals, and the RDP privacy parameter $\rho$. In addition, the mechanism takes as input the sets of admissible, protected, and outcome attributes. We estimate all of the 2-way marginals using Private-PGM~\cite{McKennaSM19}. From there in line \ref{line:exp_2_way} we measure the L1 error of each of the 2-way marginals (a sensitivity 1 estimate of the mutual information) using the Gaussian mechanism and set each edge weight to the error of its corresponding marginal. In line \ref{line:queue} we create a priority queue where we will store partial \mst s which will be sorted by their current weight in ascending order. 
In lines \ref{line:max_measure}- \ref{line:weight} we set the weight of each edge to be the maximum measurement minus the measurement for that pair. This ensures that the attribute pairs with the highest mutual information have the lowest weight and reduces the problem to that of finding the lowest weight tree. This way the partial tree with the lowest weight will always remain on the top of the queue. In order to seed the queue in line \ref{line:graph}, we add the partial tree where no nodes are connected with weight $0$. From there (lines \ref{line:start_while}-\ref{line:end_while}) at each time step we take the top partial tree from the queue and add to the queue all the possible trees with one additional edge that does not violate \cref{thrm:bayes_path}. We continue this process until the first complete \mst\ is on the top of the queue. Since the queue is sorted by weight and adding any edge only increases the weight, the first complete \mst\ to be found, by definition, must be the minimum weight tree.

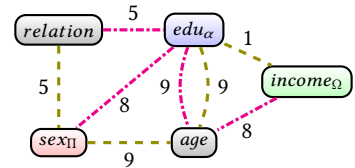
\begin{figure}

\vspace{-2mm}
\begin{tikzpicture}[scale=0.6,snode/.style = {shape=rectangle, rounded corners, draw, align=center, top color=white, bottom color=blue!20},
outcome/.style={shape=rectangle, rounded corners,thick,draw,, top color=white, bottom color=green!25}]
\begin{scope}[every node/.style={shape=rectangle, rounded corners,thick,draw,, top color=white, bottom color=blue!25}, 
other/.style={shape=rectangle, rounded corners,thick,draw, top color=white, bottom color=red!25},
none/.style={shape=rectangle, rounded corners,thick,draw,, top color=white, bottom color=light-gray}]
    \node (Q) at (3,5.5) {\footnotesize $edu_\alpha$};
    \node[other] (A) at (0,3) {\footnotesize $sex_\Pi$};
    \node[none] (B) at (0,5.5) {\footnotesize $relation$};
    \node[none] (F) at (3,3) {\footnotesize $age$};
    \node[outcome] (Q2) at (5.5,4.35) {\footnotesize $income_\Omega$};
\end{scope}

\begin{scope}[>={Stealth},
            standard/.style={draw=black, very thick}, exp/.style={draw=magenta, very thick, densely dashdotted},
              greedy/.style={draw=olive, very thick, dashed}]

    \path [-,greedy] (B) edge node[left] {5} (A);
    \path [-,exp,bend right=20] (Q) edge node[left] {9} (F);
    \path [-,greedy,bend left=20] (Q) edge node[right] {9} (F);
    \path [-,exp] (B) edge node[above] {5} (Q);
    \path [-,exp] (Q2) edge node[below] {8} (F);
    \path [-,greedy] (A) edge node[below] {9} (F);
    \path [-,greedy] (Q) edge node[above] {1} (Q2);
    \path [-,exp] (Q) edge node[below] {8} (A);
\end{scope}

\end{tikzpicture}

    \caption{Outputs of Exponential-\sys\ (magenta dashed-dotted edges) and Greedy-\sys\ (olive dashed edges).}
    \label{fig:counterexample}
\end{figure}

\begin{myexp}
Consider the graph in \cref{fig:counterexample} on the attributes of the Adult dataset. 
The magenta dashed-dotted edges show the output of Exponential-\sys\ in the absence of privacy noise. Exponential-\sys\ iterates through all permutations of partial \mst s, checking those with high edge weights. The result (in the absence of noise) is the optimal \mst\ 
with a total weight of 30.
\end{myexp}

\paratitle{Correctness and optimality condition}
We now detail the guarantees of Algorithm \ref{algo:fair_mst_2}. 
\revc{
\begin{lemma}\label{lemma:termination}
Algorithm \ref{algo:fair_mst_2} always terminates at line \ref{l:ret-tree}.
\end{lemma}

\begin{proof}
Algorithm \ref{algo:fair_mst_2} iterates through all possible \mst s  as it considers all $|\gA|^{|\gA|}$ spanning trees in lines \ref{l:inner-for} and \ref{line:end_while}, removing those which do not satisfy \cref{def:fair-mst}. 
Since a fair \mst\ always exists (see \cref{prop:existance}), Algorithm \ref{algo:fair_mst_2} will find it and will terminate at line \ref{l:ret-tree}.
\end{proof}
}
\begin{proposition} \label{prop:exp_rdp}
Given a private database $D$ over a set of attributes $\gA$, a set of admissible attributes $\admissible \subseteq \gA$, a set of protected attributes $\prot \subseteq \gA$, outcome attribute $\outcome \subseteq \gA$, measurements of 1-way marginals of $\gA$ over $D$, $log$, and a privacy parameter $\rho$, the following holds:
\begin{enumerate}
    \item Algorithm \ref{algo:fair_mst_2} satisfies $(\alpha, \rho \alpha)-RDP$ for all $\alpha \geq 1$. 
    \item Algorithm \ref{algo:fair_mst_2} outputs a fair \mst.
\end{enumerate}
\end{proposition}

While Algorithm \ref{algo:fair_mst_2} always outputs a fair \mst\ in all cases, it is only guaranteed to maximize the mutual information in the case without any privacy preserving noise. Adding privacy preserving noise to the optimal non-private mechanism does not always result in the optimal private mechanism. %

\begin{proposition}\label{prop:exp_optimal}
In the absence of privacy preserving noise (as $\rho \to \infty$), Algorithm \ref{algo:fair_mst_2} outputs the fair \mst \ with maximum mutual information.
\end{proposition}

\paratitle{Complexity and privacy budget discussion}
Algorithm \ref{algo:fair_mst_2} can be both slow and privacy intensive. First, there is an exponential number of possible spanning trees and, in the worst case, considering all of them is prohibitively expensive, resulting in the following time complexity. 

\begin{proposition}\label{prop:exp_time}
The time complexity of Algorithm \ref{algo:fair_mst_2} is $O \left(|\gA|^{|\gA| -2}\right)$.
\end{proposition}

In terms of privacy budget, since there can be a possible exponential number of additions to the partial solution queue using the exponential mechanism at each time step is not practical. Instead, the mechanism measures all of the edges using the Gaussian mechanism and repeatedly uses those values when necessary. While better this still requires a larger split of the privacy budget than the greedy solution. 
Specifically, there are $\frac{|\gA|(|\gA| -1)}{2}$ calls to the Gaussian mechanism in the algorithm. \reva{The repeated calls to the Gaussian mechanism requires that the privacy budget be split into smaller parts and as such results in higher noise in these measurements. This often leads to poor running time performance which we measure in \cref{sec:experiments}.}

\subsection{Greedy Algorithm}
\label{sec:greedy}
We now introduce an alternative solution. We slightly change the selection step of MST \cite{RyanMnist} to only consider a subset of possible edges. This will always result in a fair \mst\ but may not be optimal in terms of mutual information. However, this solution runs in linear time with respect to the number of attributes. Furthermore, we show that this solution is optimal in certain special cases.

 \begin{algorithm}[t]
\SetAlgoLined

\caption{Greedy-\sys}
\label{algo:fair_mst_1}
\SetKwInOut{Input}{input}\SetKwInOut{Output}{output}
\Input{D, 
 A, 
P, 
 O, 
 $log$, 
 $\rho$ \tcp*{Defined in Algorithm \ref{algo:fair_mst_2}}}
\Output{Set of $(i,j)$ pairs to measure $\gC$}
Use Private-PGM \cite{McKennaSM19} to estimate all 2-way marginals $\bar{M}_{i,j}$ from $log$\label{line:all-2-way}\\
Compute $L_1$ error between the estimated 2-way marginal and real 2-way marginal for all $i,j$ pairs. $q_{i,j} = \|M_{i,j}(D) -\bar{M}_{i,j}\|$ (an approximation of mutual information) \label{line:greedy_2_way}\\
Let $G = (\gA,\gC)$ be the graph where attributes are vertices and edges are pairs of attributes. \\
\tcc{Ensure a fair \mst\ (\cref{def:fair-mst})}
\textcolor{blue}{Remove the edges $(o,i)$ where $o \in O$ and $i \notin O \cup A$} \label{line:row_remove} \\
Let $r = |\gA|$ \\
Let $\epsilon= \sqrt{\frac{8 \rho}{r-1}}$\\
\For{$k=1 \text{to}\: r-1$ \label{line:start_for}}{
Let $S$ be the set of all pairs $(i,j)$ where $i$ and $j$ are in different connected components of $G$. \\
Select pair $(i,j)$ via the exponential mechanism with score function $q_{i,j}$ on set $S$ with privacy parameter $\epsilon$. \label{line:end_for}\\
add $(i,j)$ to $\gC$} 
\Return $\gC$ 

\end{algorithm}

We can adapt the greedy selection step from the MST mechanism \revc{in order to generate} a fair \mst\ by restricting the set of possible neighbors of outcome attributes. We simply restrict all the neighbors of outcome attributes to be either other outcome attributes or admissible attributes. This ensures that any path to the outcome attributes must pass through at least one admissible attribute. In practice, we employ the private Kruskal's algorithm (described in \cite{RyanMnist}) for finding a \mst\ but delete all the edges from outcome attributes to attributes that are neither outcomes nor admissible. This algorithm is shown in full in Algorithm \ref{algo:fair_mst_1}. The blue line shows the modification to the original algorithm. \par 
 Algorithm \ref{algo:fair_mst_1} takes as input the database, the measured one-way marginals and the RDP privacy parameter $\rho$. Additionally, it takes in as input the set of admissible and protected variables. We then construct a \mst\ over a complete graph where the nodes are attributes and the edges represent 2-way marginals between attributes and their weights are low sensitivity estimates of the mutual information between the two. In order to get this estimation in line \ref{line:greedy_2_way} we first estimate the 2-way marginals from the 1-way marginals using Private-PGM \cite{McKennaSM19} then set \revb{each edge's} weights to be the L1 error of those estimates. We then do the private version of Kruskal's algorithm. In line \ref{line:row_remove}, we first remove all the edges from outcome attributes to inadmissible attributes as adding any of these nodes would result in a violation of justifiable fairness. Then at each time step (lines \ref{line:start_for}-\ref{line:end_for}) we select an edge to add to the partial maximum spanning tree from the set of edges that would connect two disjoint connected components. To maintain privacy, this selection is done via the exponential mechanism. After $|\gA|-1$ rounds of this process, the resulting edges will form a private estimation of a 
 spanning tree over this graph. Once this MST is selected, 
 the distribution measured directly using the Gaussian mechanism and are then used to generate the synthetic data.
\begin{myexp}
Reconsider the graph in \cref{fig:counterexample}.
The olive dashed edges show the output of Greedy-\sys\ in the absence of privacy noise. The mechanism greedily takes the highest weight edge so long as that edge is not between an attribute in $\outcome$ and an attribute not in $\outcome \cup \admissible$. In this example it starts by taking the edge between sex and age, then takes the edge between age and education followed by the edge between sex and relation and finally the edge between education and income. This results in a final edge weight of $24$ which is slightly less than the optimal $30$ however it still results in a fair \mst. 
\end{myexp}

The proof that Algorithm \ref{algo:fair_mst_1} satisfies RDP is identical to the proof that the original MST algorithm satisfies DP \cite{RyanMnist}. 
Furthermore, since Algorithm \ref{algo:fair_mst_1} only allows attributes in $\outcome$ to be adjacent to attributes in $\admissible$ regardless of the sampling order each $\outcome$ is independent of all other nodes given the attributes in $\admissible$ and, therefore, generates a fair \mst. As such, we get the following.

\begin{proposition} \label{prop:greedy_rdp}
Given a private database $D$, a set of admissible attributes \admissible, a set of protected attributes \prot, outcome attribute \outcome, measurements of 1-way marginals of the attributes of $D$, $log$, and a privacy parameter $\rho$, the following holds:
\begin{enumerate}
    \item Algorithm \ref{algo:fair_mst_1} satisfies $(\alpha, \rho \alpha)-RDP$ for all $\alpha \geq 1$.
    \item Algorithm \ref{algo:fair_mst_1} outputs a fair \mst.
\end{enumerate}
\end{proposition}

\paratitle{Optimality for the saturated case}
Algorithm \ref{algo:fair_mst_1} is not optimal (in the absence of noise) in the general case as there are cases where the greedy solution results in a fair \mst \ with sub-optimal total mutual information. There is, however, a special case for which Algorithm \ref{algo:fair_mst_1} is optimal. When all attributes are either admissible protected or outcomes (i.e., $\admissible \cupdot \outcome \cupdot \prot = \gA$), we say that the problem instance is {\em saturated}.  The saturated case corresponds to the cases where the user has complete knowledge as to which attributes may or may not be used for decision making and there is no ambiguity among the attributes. In the saturated case, we note that in any fair \mst, all edges from outcome attributes must either come from other outcome attributes or admissible attributes.

\begin{proposition}\label{prop:fair_parents}
Given a database $D$ over attributes $\gA$, and disjoint sets $\prot,\admissible,\outcome \subseteq \gA$ representing the protected, admissible, and outcome attributes respectively, a \mst\ $T$ whose node set is $\gA$ is fair if all the neighbors of nodes in the set $\outcome$ are in the set $\admissible \cup \outcome$. If the problem is saturated the converse is also true. 
\end{proposition}

Thus, when the problem is saturated, finding the optimal fair \mst\ is equivalent to finding the MST on the subgraph containing no edges from attributes in $\prot$ to attributes not in $\prot$ or $\admissible$. Therefore, using a traditional MST algorithm such as Kruskal's algorithm \cite{Kruskal} on this subgraph results in an optimal \mst. Thus, using Algorithm \ref{algo:fair_mst_1} yields an optimal fair \mst.

\paratitle{Complexity and privacy budget discussion}
Algorithm \ref{algo:fair_mst_1} is efficient with respect to both privacy budget and computation time. Unlike Algorithm \ref{algo:fair_mst_2} which uses a quadratic number of calls to the Gaussian Mechanism, Algorithm \ref{algo:fair_mst_1} calls the exponential mechanism only a linear number of times.  This results in significantly more privacy budget being allocated to each individual call of the exponential mechanism resulting in lower noise.
Algorithm \ref{algo:fair_mst_1} also has a significantly improved run time. Since Algorithm \ref{algo:fair_mst_1} only selects $|\gA|-1$ edges using the exponential mechanism it only requires a linear time to run, in addition to the quadratic time for computing the 2-way marginals (lines \ref{line:all-2-way},\ref{line:greedy_2_way}). This is a dramatic improvement over Algorithm \ref{algo:fair_mst_2} which is exponential. %

\begin{proposition}\label{prop:greedy_time}
The time complexity of Algorithm \ref{algo:fair_mst_1} is $O \left(|\gA|^2\right)$.
\end{proposition}

\section{Experiments}\label{sec:experiments}

We present experiments that evaluate the efficacy of the proposed mechanisms. In particular, we explore the following main questions.
\begin{enumerate}[topsep=0pt,itemsep=-1ex,partopsep=1ex,parsep=1ex, leftmargin=*]

    \item \textbf{Q1}: What is the error overhead incurred by Algorithms \ref{algo:fair_mst_2} and \ref{algo:fair_mst_1} compared to MST~\cite{RyanMnist}?
    \item \textbf{Q2:} How effective are Algorithms \ref{algo:fair_mst_2} and \ref{algo:fair_mst_1} at reducing unfairness in downstream classification tasks?

    \item \textbf{Q3:} What is the performance of Algorithms \ref{algo:fair_mst_2} and \ref{algo:fair_mst_1} compared to the baseline that does not consider fairness?

\end{enumerate}

\paratitle{Implementation Details}
We have implemented our system in Python 3.8 based on the existing MST approach~\cite{RyanMnist}. All values shown are averaged over $10$ instances of synthetic data.

\subsection{Experimental Setup}\label{sec:setup}
We next detail the settings of the experiments including the datasets used, the examined approaches, and the employed measures. 

\paratitle{Datasets}
We test our mechanisms on \revc{three} datasets, each composed of a single table. 
\begin{itemize}[topsep=0pt,itemsep=-1ex,partopsep=1ex,parsep=1ex, leftmargin=*]
    \item {\bf Adult \cite{adult}:} This dataset contains 14 attributes and 48,843 tuples. It contains individual' census data. The objective is to predict if an individual's income is above or below 50K.
    \item {\bf Compas \cite{compas_data}:} This dataset contains 8 attributes and 6173 tuples. It contains the criminal history, jail and prison time, demographics and Compas risk scores for defendants from Broward County from 2013 and 2014. The objective is to predict if an individual will commit a crime again within 2 years.
    \item \revc{\textbf{Census KDD \cite{adult,SynthBenchmark,FairData}}: This dataset contains 41 attributes and 299,285 tuples. It contains information from the 1994 and 1995 Population Survey from the US census. The objective is to predict if an individual's income is above or below 50K.}

\end{itemize}
To prepare the data for each task, we transformed categorical data into equivalent numerical data by mapping each domain value to a unique integer. \revc{Continuous data is instead treated as discrete categorical data with each unique value being its own category (see discussion on discretization in \cref{sec:extensions})}. We have listed the set of protected, admissible, and outcome variables in \cref{tab:attributes}. 

\paratitle{Baselines and examined approaches}
For baselines, we compare against the original versions of MST \cite{RyanMnist}. For our mechanisms, we use adapted versions of each of these baselines, Exponential-\sys\ (Algorithm \ref{algo:fair_mst_2}) and Greedy-\sys\ (Algorithm \ref{algo:fair_mst_1}).

Each mechanism is used with 3 different settings of $\epsilon = [0.1,1,10]$.

We found the optimal RDP parameters and converted back to traditional DP using \cref{thrm:RDPtoDP}. Each mechanism is set to generate a dataset of the same size as the original dataset. All measurements are done over 10 samples of synthetic data for each mechanism. 

We generated learned models using Tensorflow \cite{tensorflow} for MLP models and scikit-learn \cite{scikit-learn} for linear regression as well as random forests. Each model was trained using one set of synthetic data and was tested against the underlying true data.

\begin{table}[t]
\caption{Dataset attributes division.}
\begin{small}
\begin{tabularx}{\columnwidth}{|c|c|X|}
    \hline
    {\bf Dataset} & {\bf Division} & {\bf Attributes} \\
    \hline
    \multirow{4}{*}{Adult} & Protected  & Sex, race, native country  \\
    \cline{2-3}
    & Outcome & Income\\
    \cline{2-3}
    & \multirow{2}{*}{Admissible}  & Workclass, education, occupation, capital-gain, capital-loss, and hours per week\\
    \hline\hline
    \multirow{3}{*}{Compas} & Protected  & Sex, race  \\
    \cline{2-3}
    & Outcome & Two Year Recidivism\\
    \cline{2-3}
    & \multirow{1}{*}{Admissible}  & Number of Priors, misdemeanor/felony\\
    \hline\hline
    \multirow{3}{*}{Census KDD} & Protected  & Sex, race  \\
    \cline{2-3}
    & Outcome & Income\\
    \cline{2-3}
    & \multirow{1}{*}{Admissible}  & Workclass, Industry, Occupation, Education $\dots$\\
    \hline
\end{tabularx}
\end{small}
\label{tab:attributes}
\vspace{-.5em}
\end{table}

\paratitle{Data quality}
In order to measure the overall quality of the data, we consider the same set of metrics as \common{past benchmarks} \cite{SynthBenchmark}. Each of these metrics are intended to measure a different quality of the data that should be preserved in the synthetic data.
\begin{enumerate}[topsep=0pt,itemsep=-1ex,partopsep=1ex,parsep=1ex, leftmargin=*]
    \item \textbf{Individual Attribute Distribution:} We measure the similarity of the 1-way marginals between the synthetic and original data. To do so, we measure the Total Variation Distance (TVD) between a vectorized version of the 1-way marginals for both the synthetic and original data. The value shown is the \reva{average} of the total variation distances for each 1-way marginal. We show the distribution over 10 samples of synthetic data. 
    \item\textbf{Pairwise Attribute Distribution:} We extend the methodology of 1-way marginals to instead measure the distributions of 2-way marginals as well. As before we measure the TVD of the 2-way marginals vectors of the original and synthetic data and show the \reva{average} over all 2-way marginals. 
    \item \textbf{Pairwise Correlation Similarity:} As in~\cite{SynthBenchmark}, we measure the Cramer's V with bias correction measure between each pair of attributes. \reva{Cramer's V, which lies in the range $[0,1]$, is a measure of correlation between two attributes with higher values signifying a larger correlation. } We measure the difference between the measures in the synthetic and original data and show the \reva{average} of the differences across all pairs of attributes, \reva{a measure we call the average correlation difference (ACD)}.  
    \item \textbf{Downstream Classification Accuracy:} Here we use the synthetic data to train several classifiers and use it to classify the original data. We report the accuracy $\frac{TP + TN}{TP+FP+FN+TN}$ for a model trained on each of the 10 output samples.
\end{enumerate}

 \begin{table}[t]
\caption{Fairness measures.}
\begin{small}
\begin{tabularx}{\columnwidth}{|l|X|}

    \hline
    DP & $ Pr(O=1|S=1) - Pr(O=1|S=0)$\\
    \hline
    TPRB  &  $   Pr(O=1|S=1, Y=1) - Pr(O=1|S=0, Y=1)$ \\
    \hline
    TNRB &  $Pr(O=0|S=1, Y=0) - Pr(O=0|S=0, Y=0)$  \\
    \hline
    CDP  & $ \mathbb{E}_{A}(Pr(O=1|S=1,A = a) - Pr(O=1|S=0, A = a)]$ \\
    \hline
    \multirow{2}{*}{CTPRB} &  $ \mathbb{E}_{A}[  Pr(O=1|S=1, Y=1,A = a) - Pr(O=1|S=0, Y=1,A = a) ]$\\
    \hline
    \multirow{2}{*}{CTNRB} & $ \mathbb{E}_{A}[ Pr(O=0|S=1, Y=0,A = a) - Pr(O=0|S=0, Y=0,A = a)]$ \\
    \hline
\end{tabularx}
\end{small}
\label{tab:fair_measures}
\end{table}

 \begin{table*}[t]
 \caption{\common{Overview of Greedy-\sys 's (Algorithm \ref{algo:fair_mst_1}) Performance against MST. All values are the percent of Greedy-\sys 's performance with respect to MST's. Data quality values use $\epsilon =1$ and ML values use the MLP model. Green cells show the favorable scenarios for Greedy-\sys.}}
  \label{tab:overview}
 \centering
 \small
 \setlength{\tabcolsep}{2pt}
 \begin{tabularx}{0.85\linewidth}{|c|X|c|c|c|c|X|c|c|c|c|c|c|X|c|}
     \hhline{~~----~------~~}
       \multicolumn{1}{c}{}&  & \multicolumn{4}{c|}{{\cellcolor{light-light-gray} \bf Quality Measures}}& & \multicolumn{6}{c|}{{\cellcolor{light-light-gray} \bf Fairness Measures}} & \multicolumn{1}{c}{}  \\
     \hhline{-~----~------~-} 
     \cellcolor{light-light-gray} {\bf Dataset} & & \cellcolor{light-light-gray} {\bf TVD 1-Way} & \cellcolor{light-light-gray} {\bf TVD 2-Way} &  \cellcolor{light-light-gray} {\bf ACD} & \cellcolor{light-light-gray} {\bf Accuracy} & & \cellcolor{light-light-gray} {\bf DP} & \cellcolor{light-light-gray} {\bf TPRB} & \cellcolor{light-light-gray} {\bf TNRB} & \cellcolor{light-light-gray} {\bf CDP}& \cellcolor{light-light-gray} {\bf CTPRB} & \cellcolor{light-light-gray} {\bf CTNRB} & &  \cellcolor{light-light-gray} {\bf Runtime}  \\
     \hhline{-~----~------~-} 
      \textbf{Adult}   & & \cellcolor{mid-red} 100.5\% & \cellcolor{mid-red} 106.0 \% & \cellcolor{mid-red}131.6 \% & \cellcolor{mid-red} 99.8\% & & \cellcolor{mid-green} 64.4\% & \cellcolor{mid-green} 67.9\% & \cellcolor{mid-green} 10.2\% & \cellcolor{mid-green} 64.8\% & \cellcolor{mid-green} 25.7\% & \cellcolor{mid-green} 11.7\% & & \cellcolor{mid-green} 92.9\%\\
     \hhline{-~----~------~-} 
      \textbf{Compas}   & & \cellcolor{mid-green} 96.5\% & \cellcolor{mid-red} 119.8\% & \cellcolor{mid-red}103.3\% & \cellcolor{mid-red} 86.2\% & & \cellcolor{mid-green} 66.2\% & \cellcolor{mid-green} 70.3\% & \cellcolor{mid-green} 73.0\% & \cellcolor{mid-green} 50.0\% & \cellcolor{mid-green} 50.5\% & \cellcolor{mid-green} 53.1\% & & \cellcolor{mid-green} 97.8\%\\
     \hhline{-~----~------~-} 

      \textbf{KDD}& & \cellcolor{mid-red} 101.4\% & \cellcolor{mid-red} 100.8\% & \cellcolor{mid-red} 104.3 \%& \cellcolor{mid-red} 99.6\% & & \cellcolor{mid-green} 62.5 \%& \cellcolor{mid-green} 81.8 \%& \cellcolor{mid-green} 24.4\% & \cellcolor{mid-green} 22.5\% & \cellcolor{mid-green} 31.8\% & \cellcolor{mid-green} 24.4\% & & \cellcolor{mid-green} 92.8\% \\
     \hhline{-~----~------~-} 
    
 \end{tabularx}
\vspace{-2mm}
 \end{table*}

\begin{figure*}[ht]
\resizebox{\textwidth}{!}{%
	\centering
	\begin{subfigure}[b]{0.35\linewidth}
		\includegraphics[width=1\textwidth]{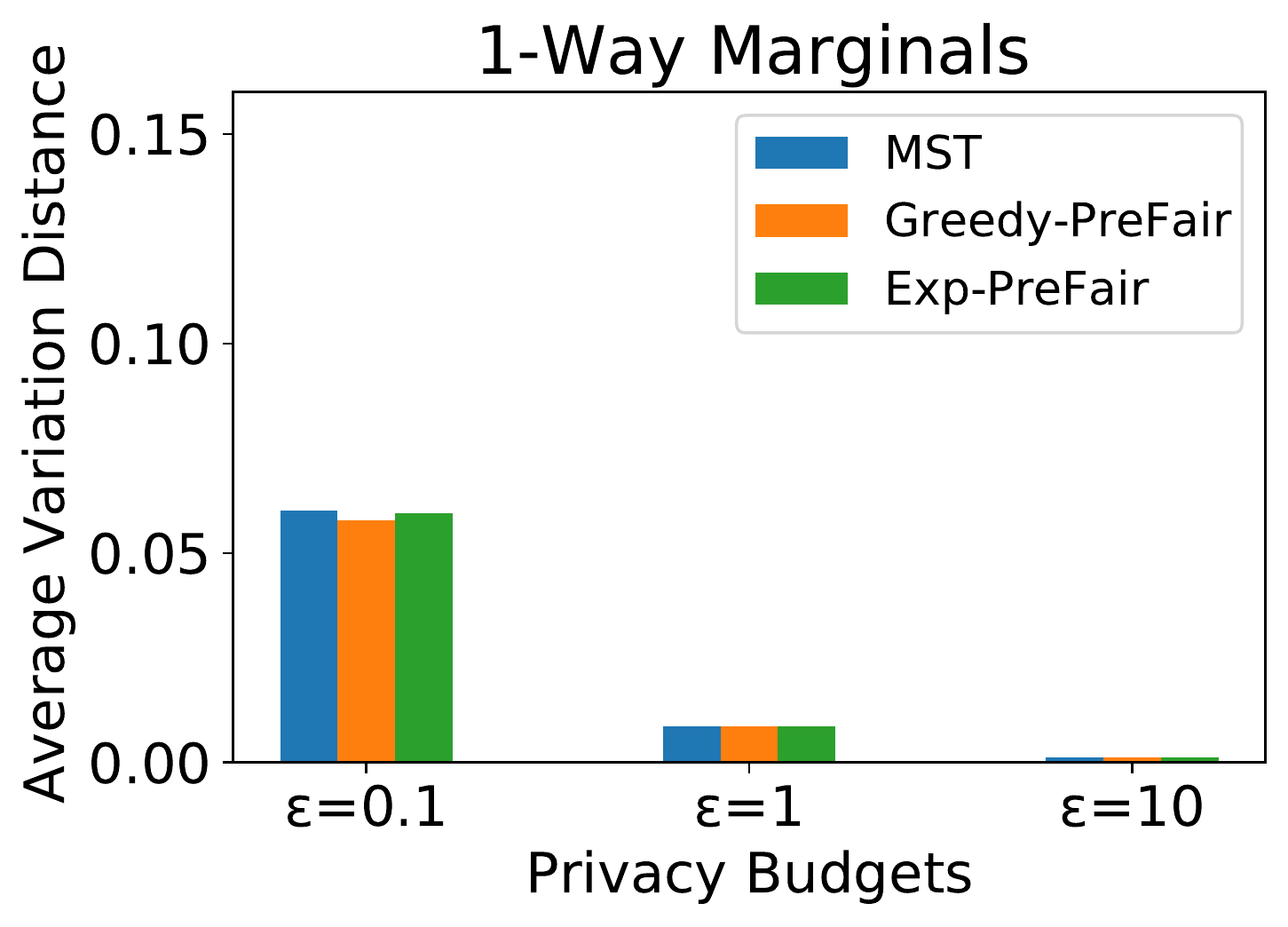}
		\caption{\reva{1-Way Marginals}}
		\label{fig:1way}
	\end{subfigure}
	\begin{subfigure}[b]{0.35\linewidth}
		\includegraphics[width=1\textwidth]{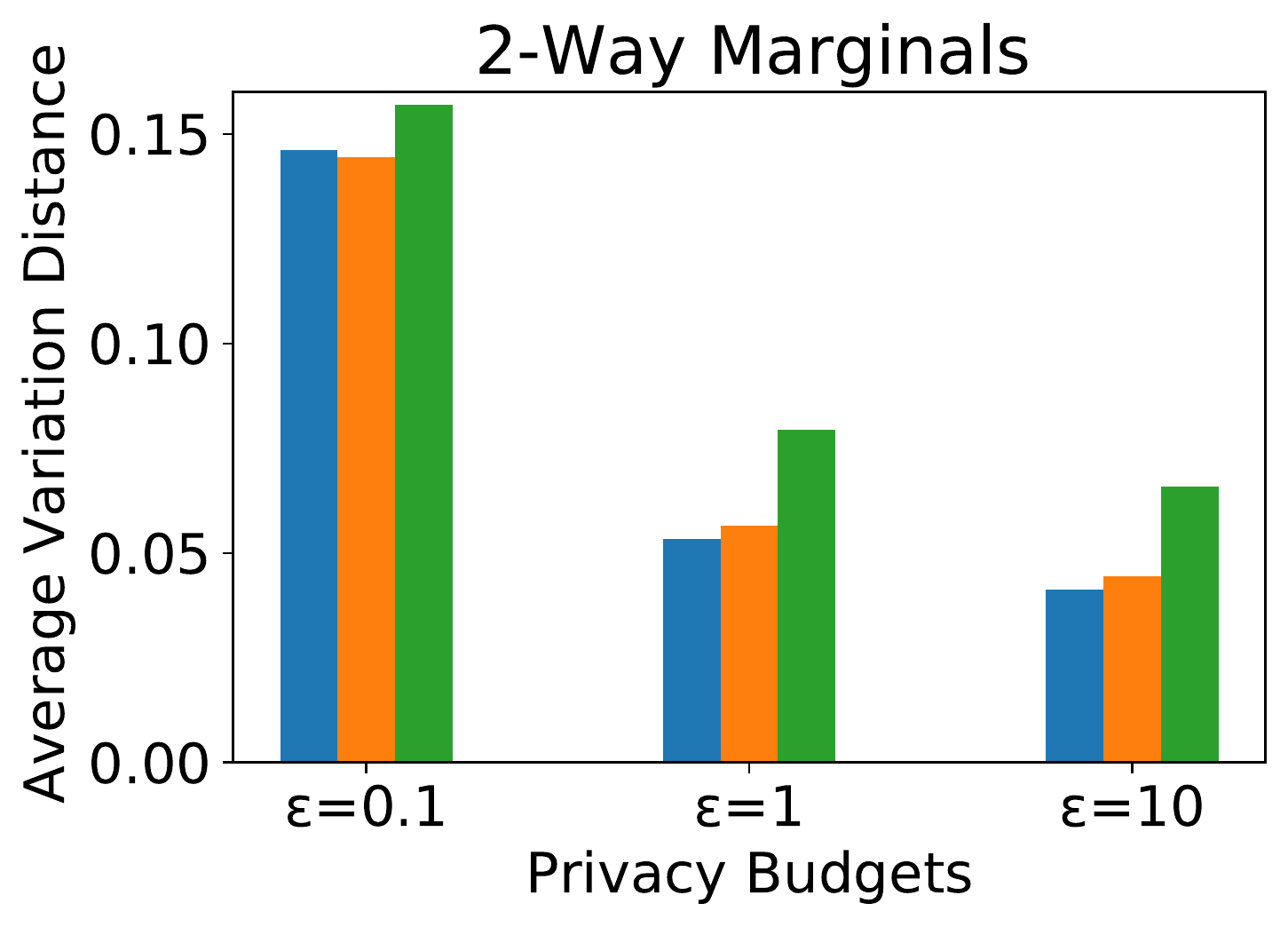}
		\caption{\reva{2-Way Marginals}}
		\label{fig:2way}
	\end{subfigure}
	\begin{subfigure}[b]{0.35\linewidth}
		\includegraphics[width=1\textwidth]{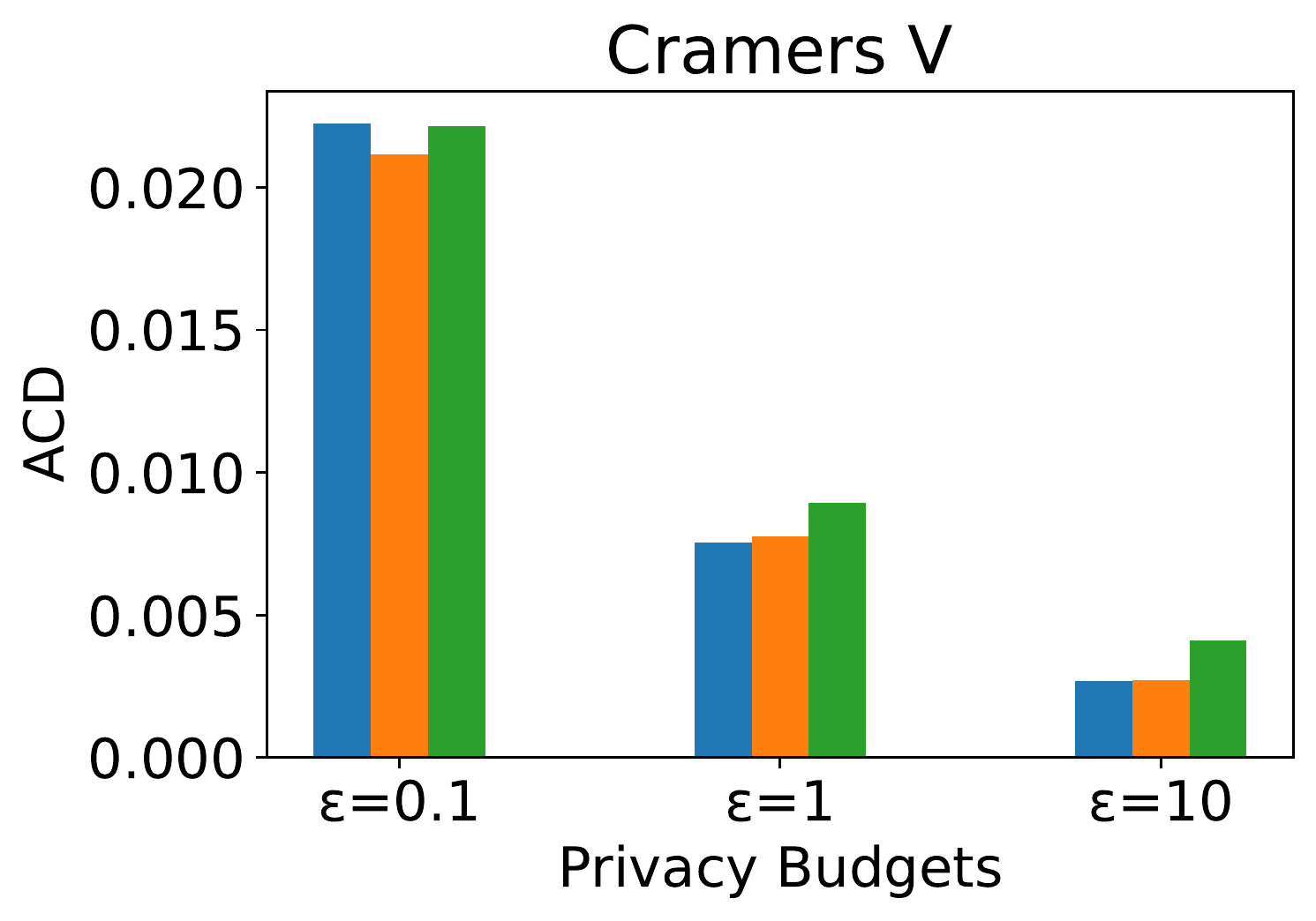}
		\caption{\reva{Cramers V Correlation}}
		\label{fig:Corr}
	\end{subfigure}
	\begin{subfigure}[b]{0.35\linewidth}
		\includegraphics[width=1\textwidth]{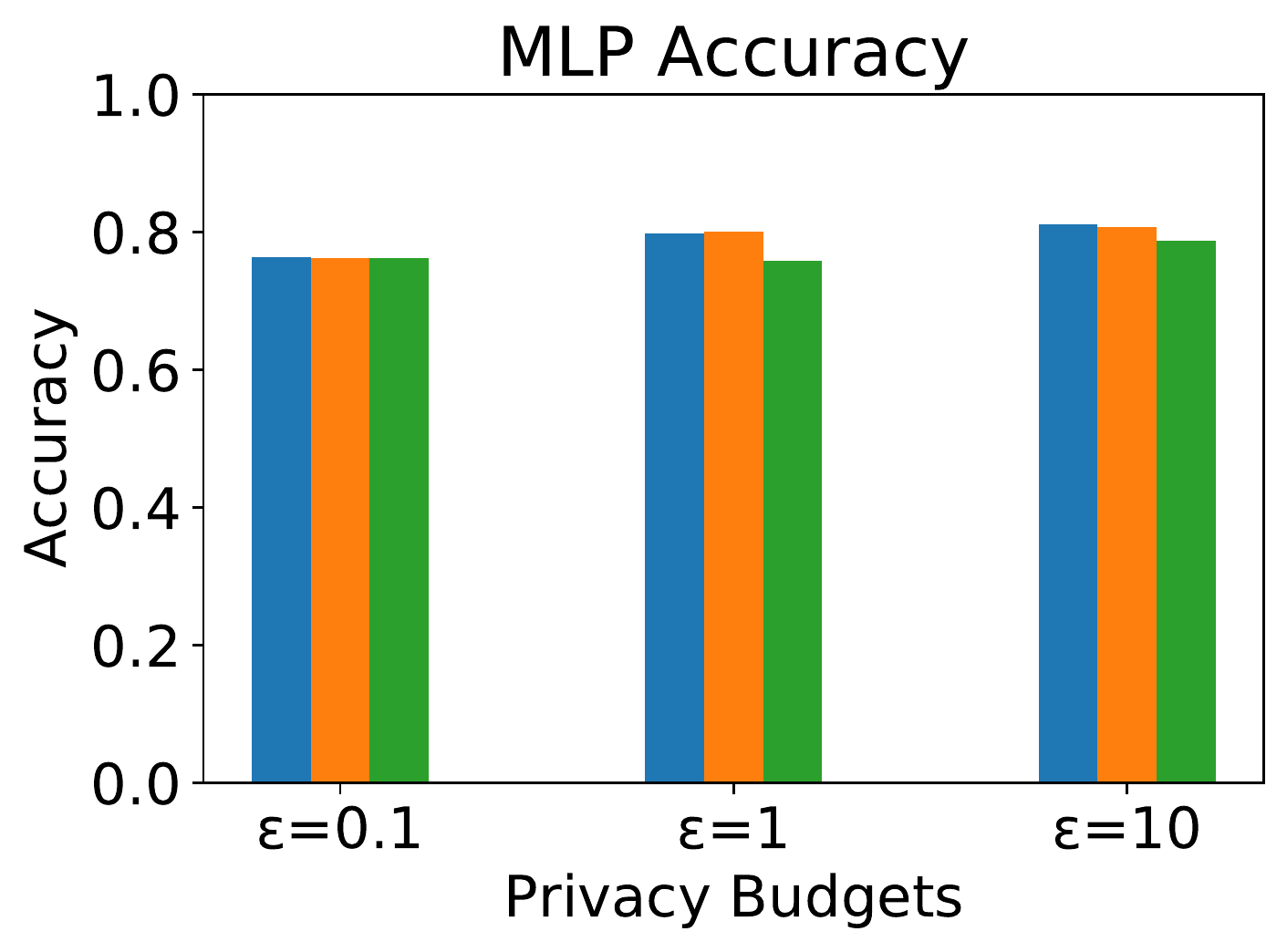}
		\caption{MLP Accuracy}
		\label{fig:MLPACC}
	\end{subfigure}
}
\caption{Data Quality Measures of the Synthetic Data For Adult.}
\label{fig:data}
\vspace{-3mm}
\end{figure*}

\begin{figure}[t]

\begin{subfigure}[b]{0.36\linewidth}
\includegraphics[width=\linewidth]{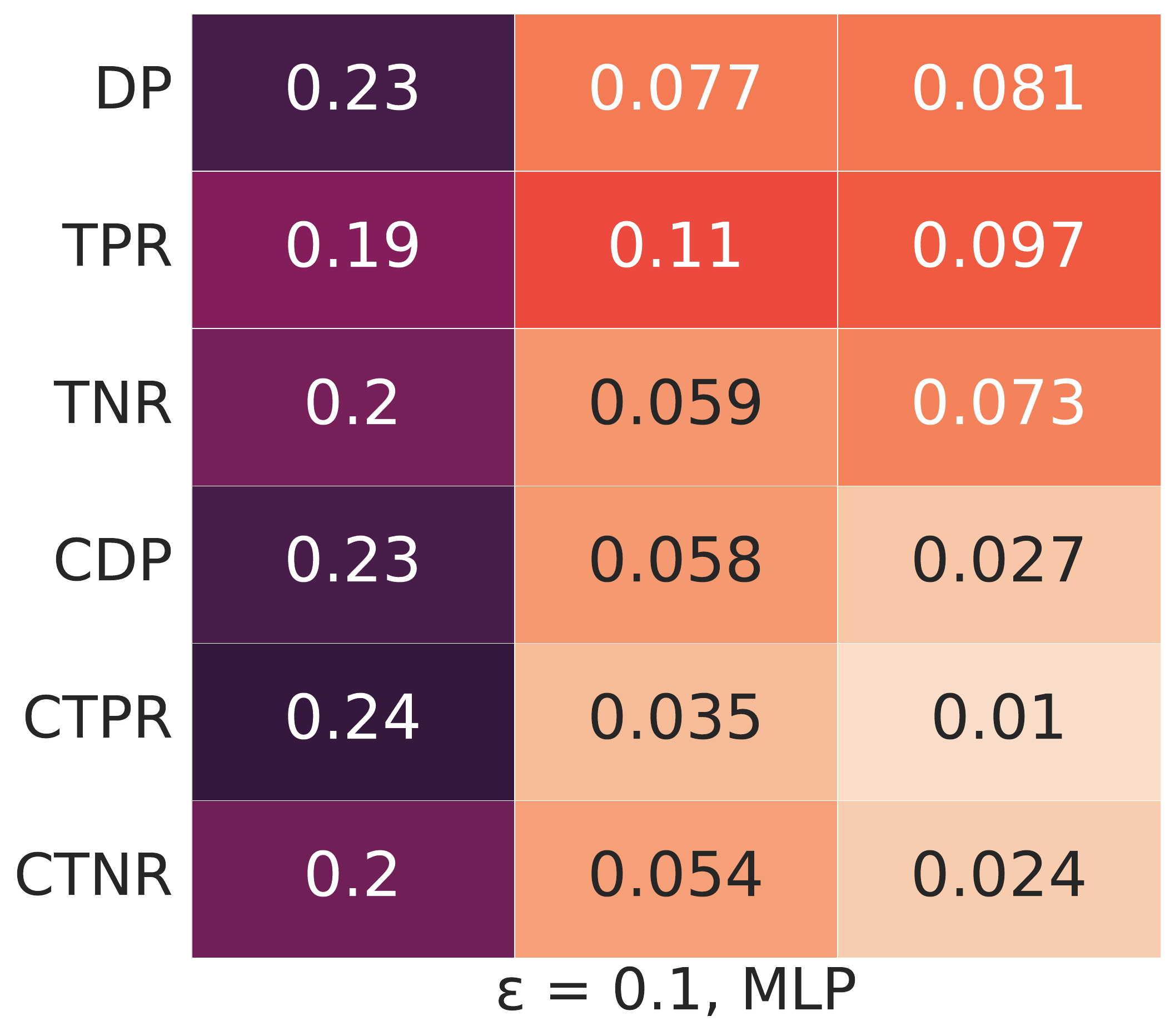}
\caption*{} \label{fig:mlp01}
\end{subfigure}\hspace*{\fill}
\begin{subfigure}[b]{0.31\linewidth}
\includegraphics[width=\linewidth]{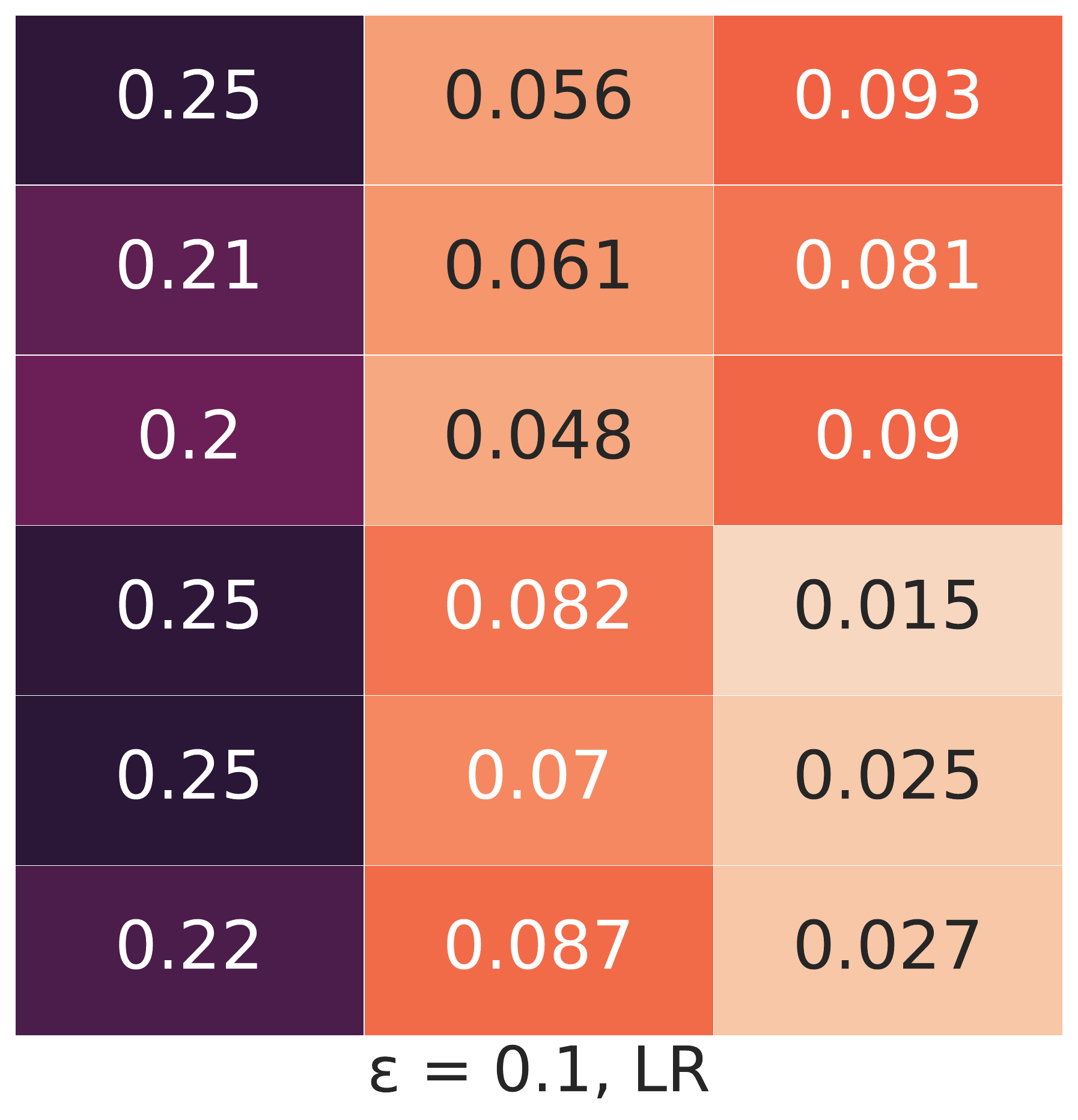}
\caption*{} \label{fig:lr01}
\end{subfigure}\hspace*{\fill}
\begin{subfigure}[b]{0.31\linewidth}
\includegraphics[width=\linewidth]{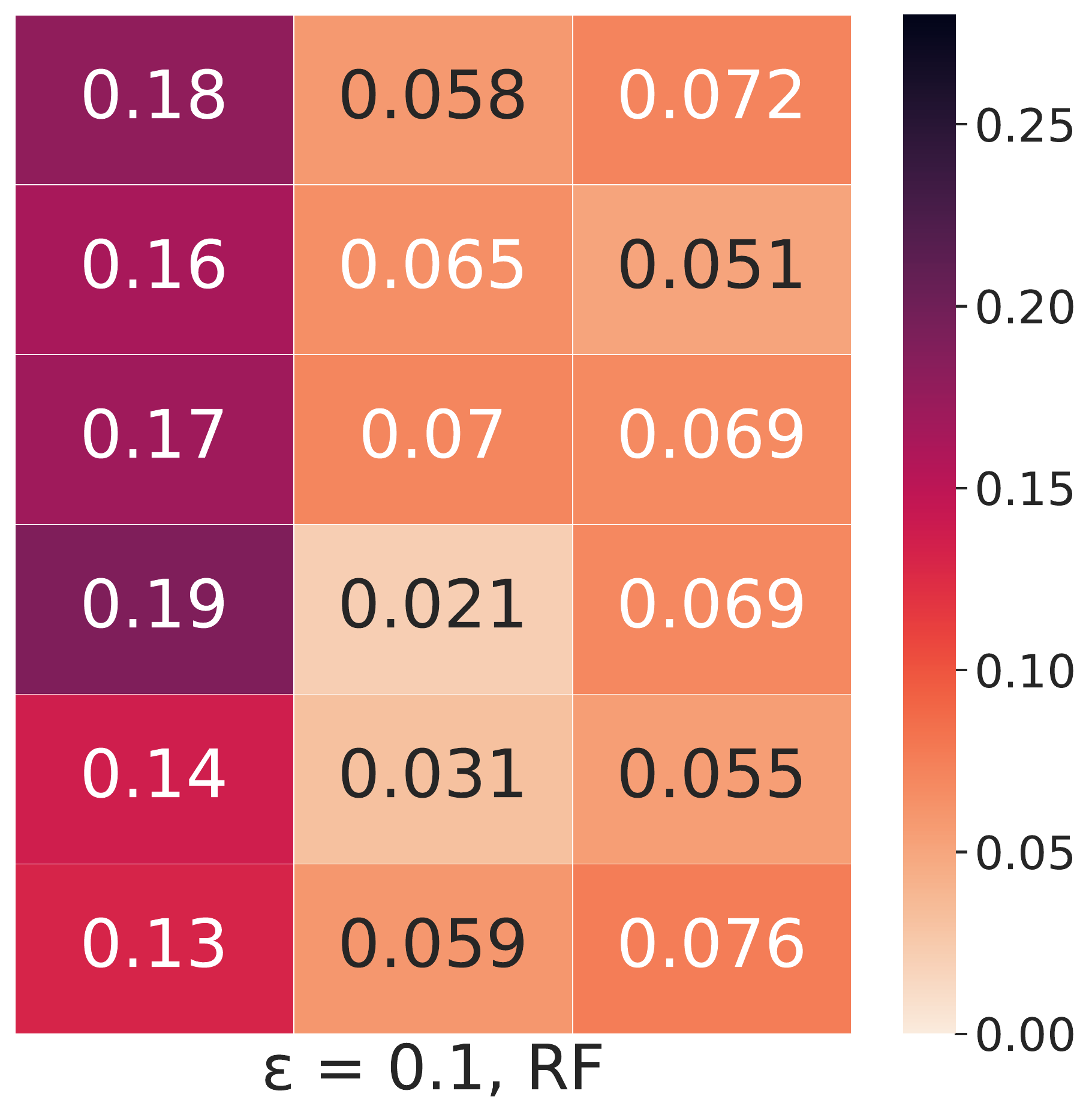}
\caption*{} \label{fig:rf01}
\end{subfigure}

\vspace{-5mm}

\begin{subfigure}[b]{0.36\linewidth}
\includegraphics[width=\linewidth]{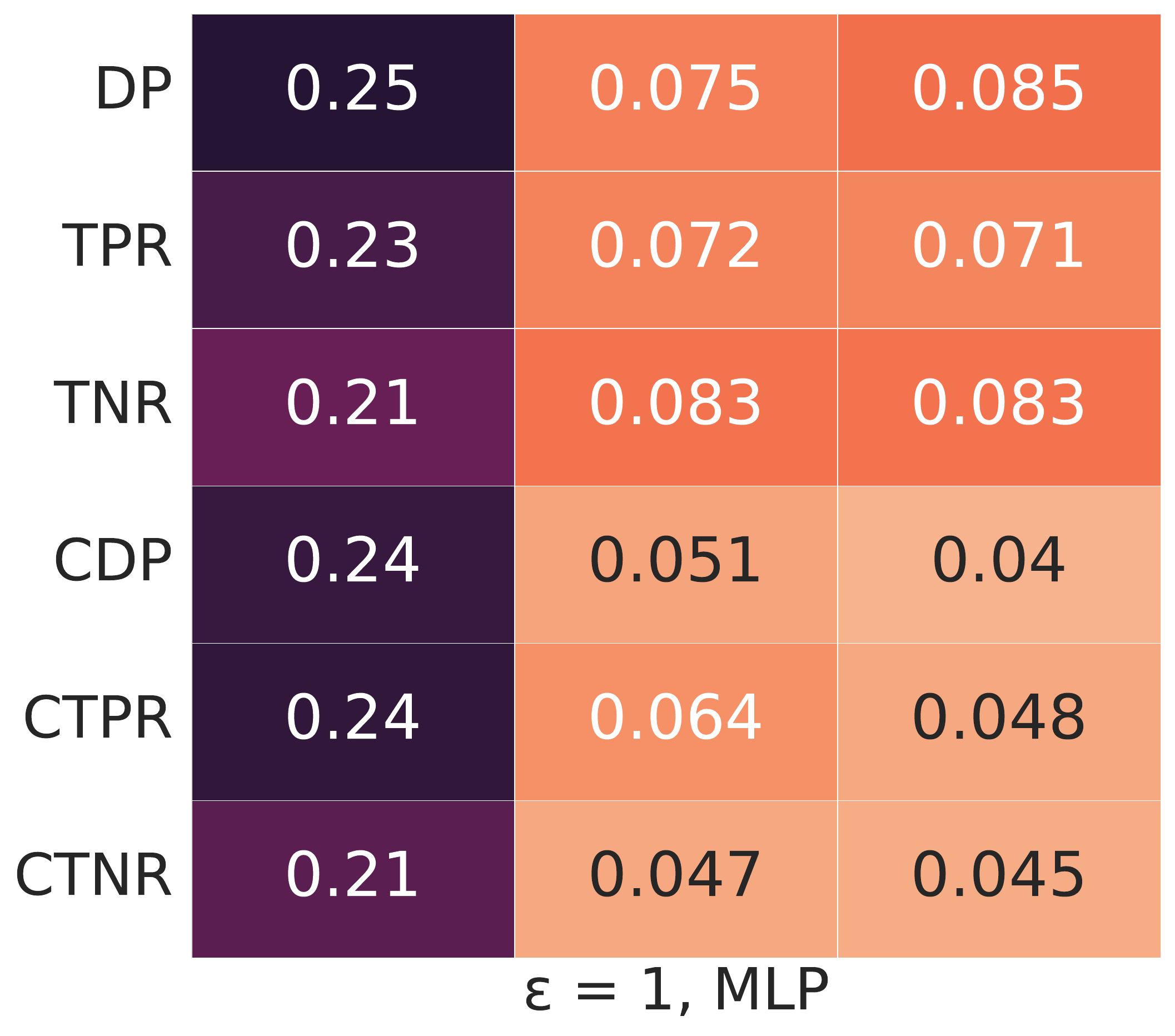}
\caption*{} \label{fig:mlp1}
\end{subfigure}\hspace*{\fill}
\begin{subfigure}[b]{0.31\linewidth}
\includegraphics[width=\linewidth]{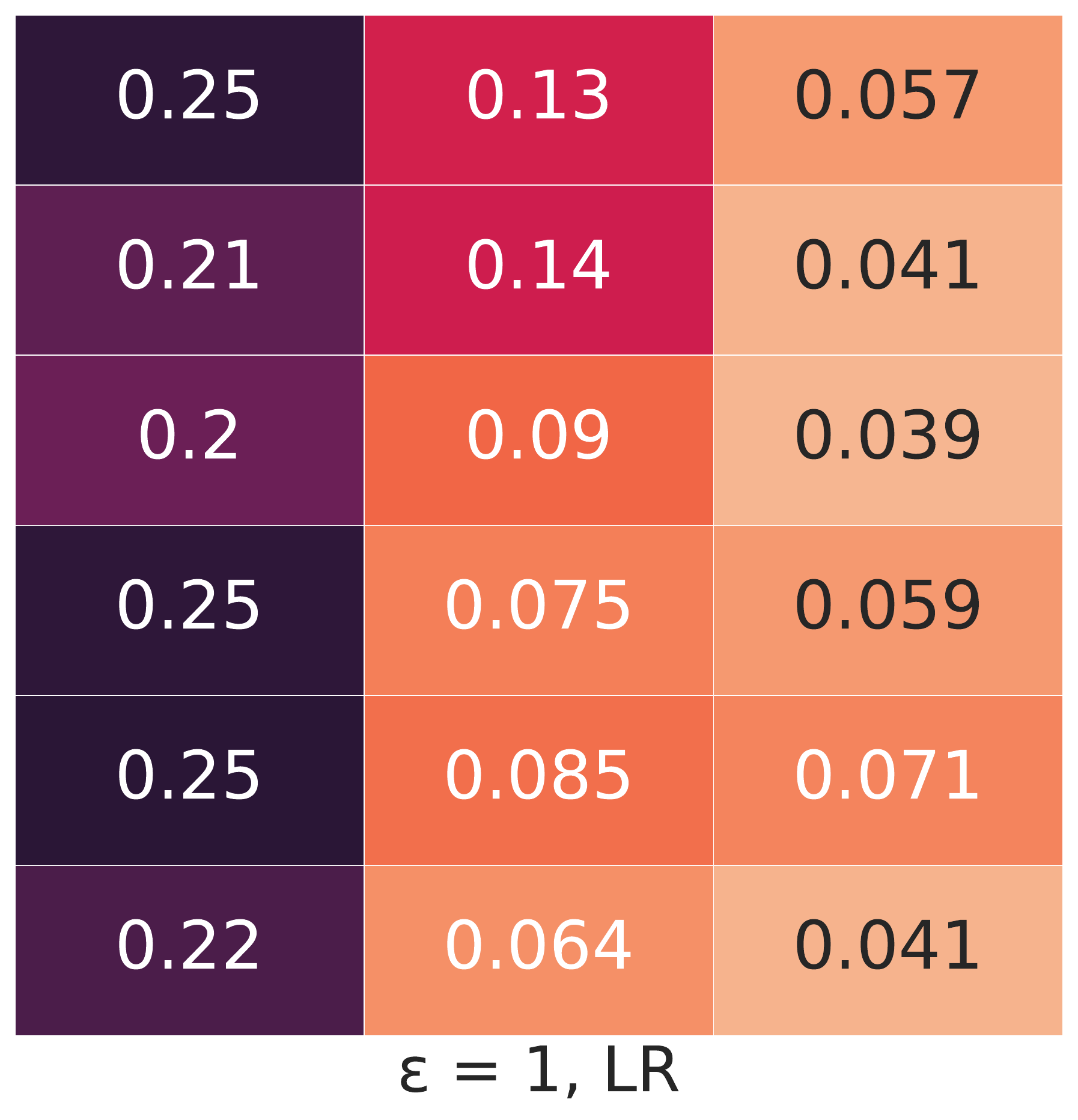}
\caption*{} \label{fig:lr1}
\end{subfigure}\hspace*{\fill}
\begin{subfigure}[b]{0.31\linewidth}
\includegraphics[width=\linewidth]{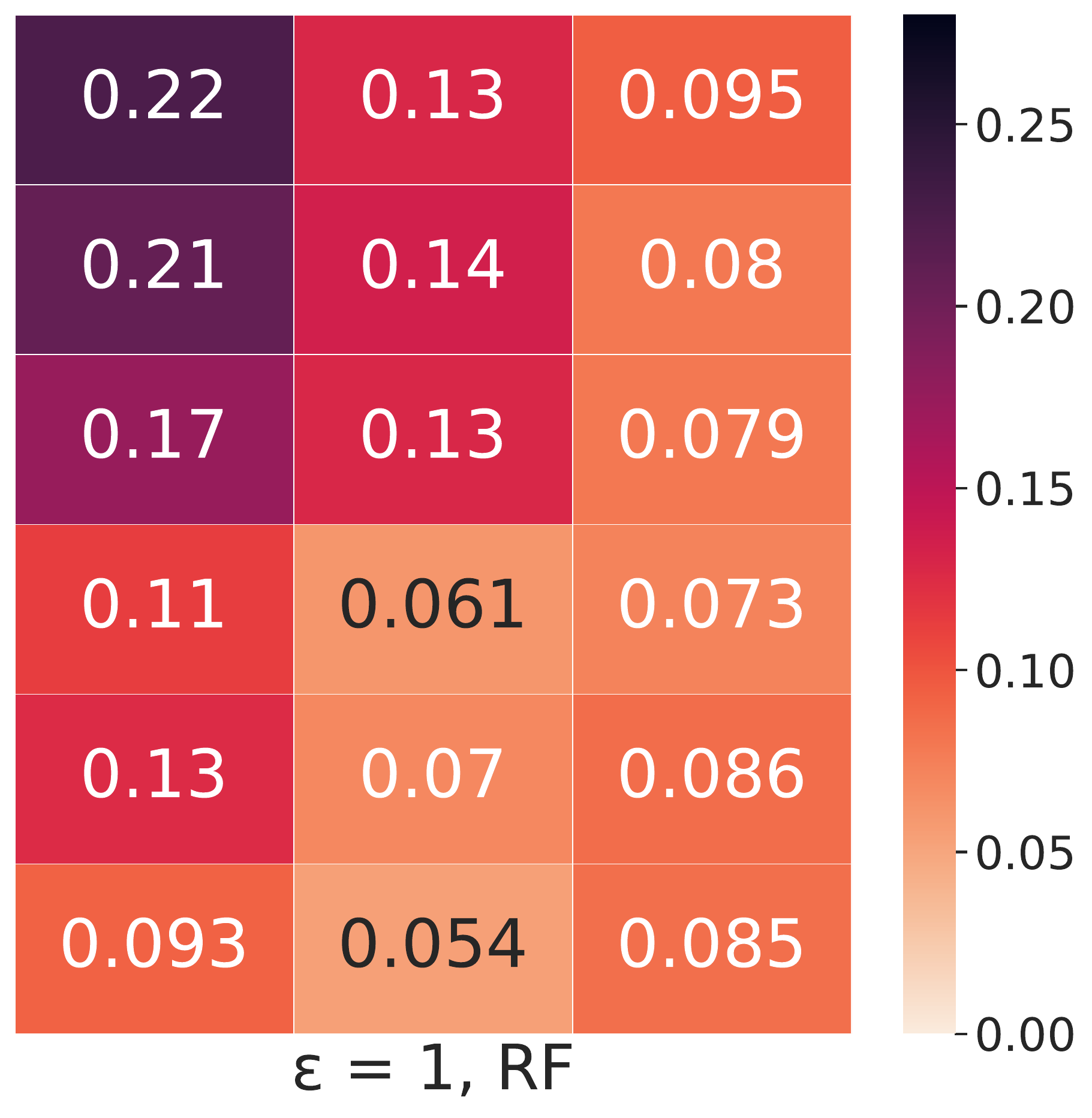}
\caption*{} \label{fig:rf1}
\end{subfigure}

\vspace{-5mm}

\begin{subfigure}[b]{0.36\linewidth}
\includegraphics[width=\linewidth]{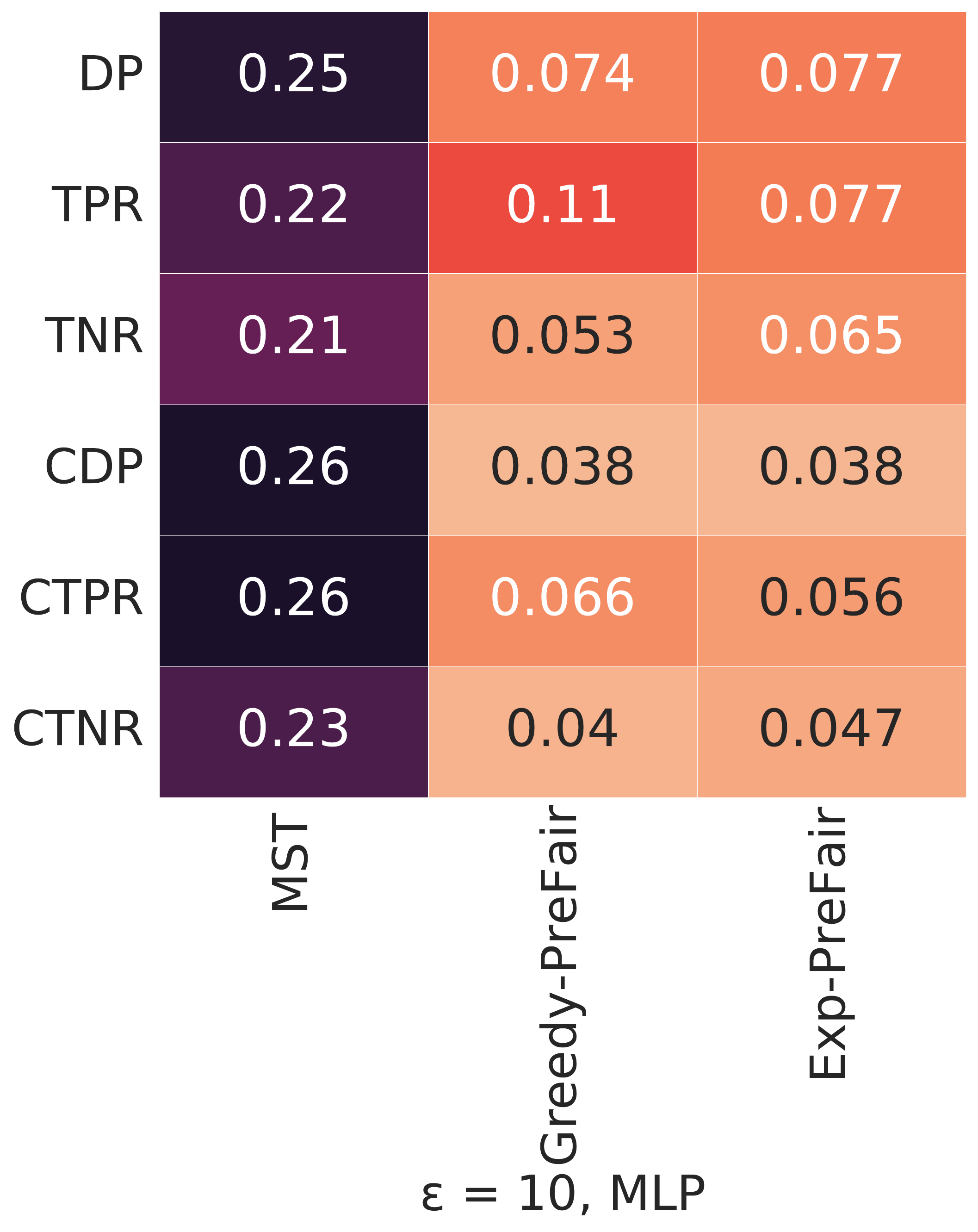}
\caption*{} \label{fig:mlp10}
\end{subfigure}\hspace*{\fill}
\begin{subfigure}[b]{0.31\linewidth}
\includegraphics[width=\linewidth]{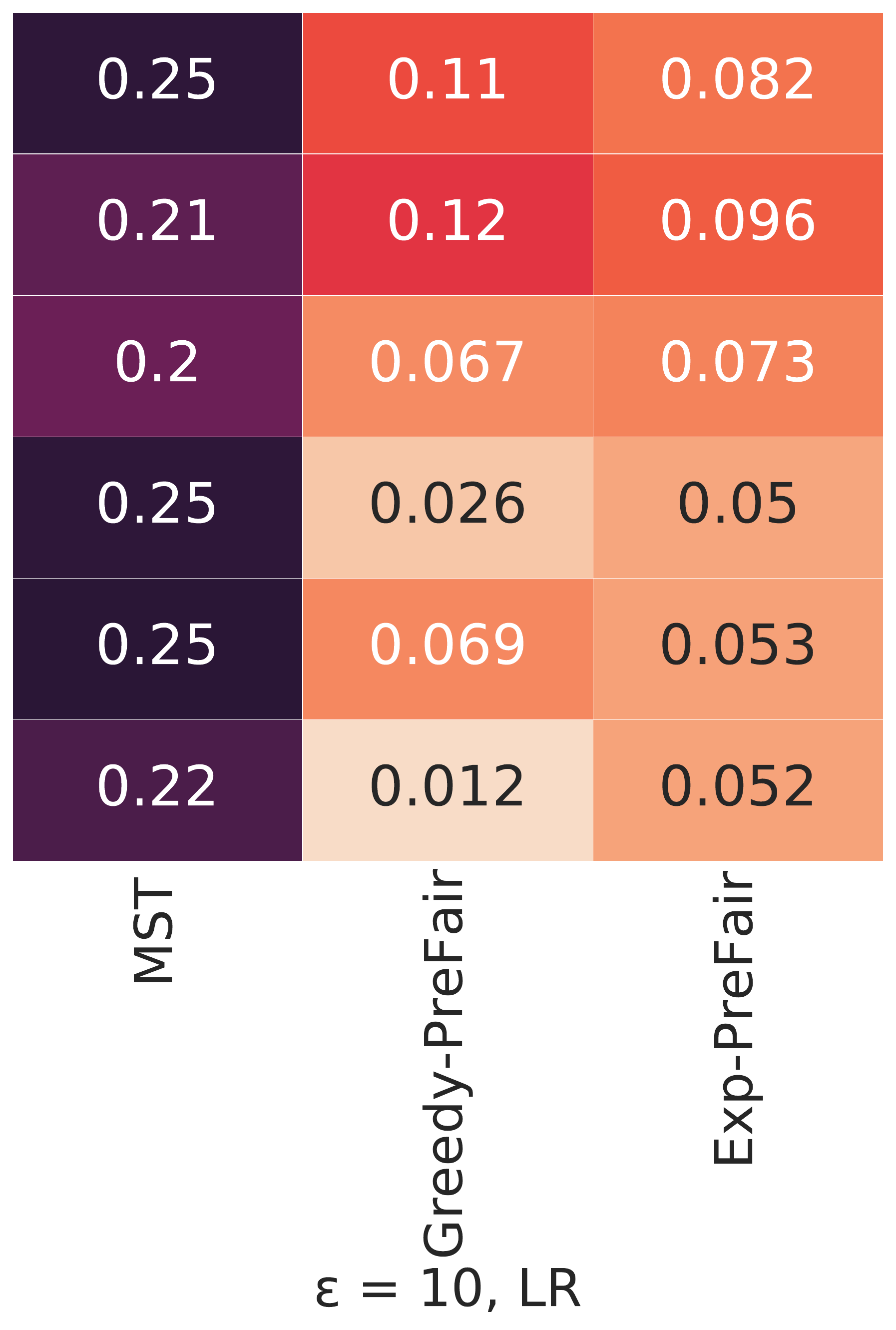}
\caption*{} \label{fig:lr10}
\end{subfigure}\hspace*{\fill}
\begin{subfigure}[b]{0.31\linewidth}
\includegraphics[width=\linewidth]{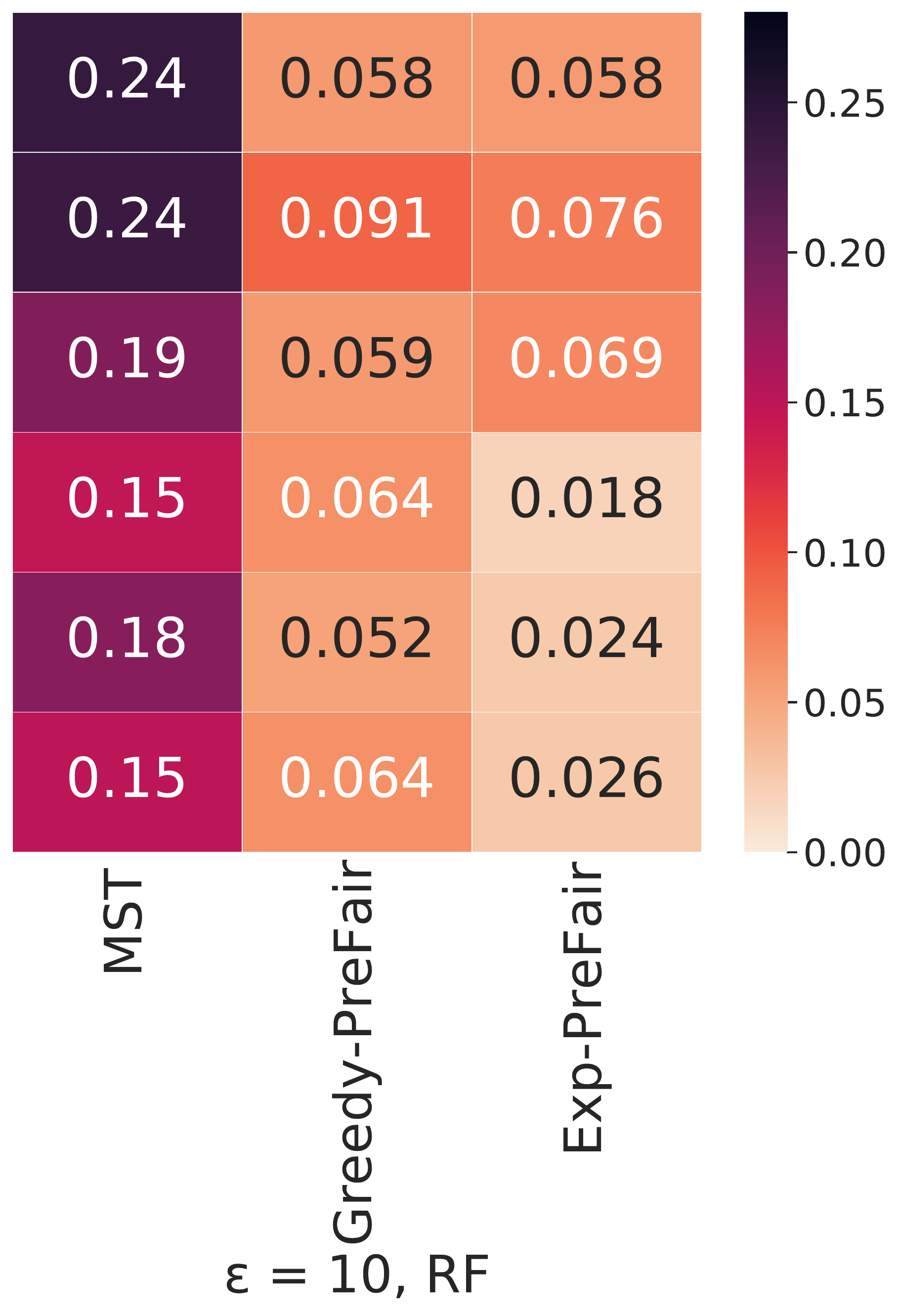}
\caption*{} \label{fig:rf10}
\end{subfigure}

\vspace{-5mm}
\caption{\reva{Fairness measurements for the Compas dataset.}
} \label{fig:fair_compas}
\end{figure}

\paratitle{Fairness metrics}
We also measure the impact that training on the fairness constrained synthetic data has on downstream classification fairness. For each of these measures, we trained a MLP, \common{Random Forrest, and Linear Regression} models on the synthetic data and measured commonly used empirical indicators of fairness. 

In order to establish our results, we employ many of the same measures used in \cite{InterFair}. 
\par 

\begin{enumerate}[topsep=0pt,itemsep=-1ex,partopsep=1ex,parsep=1ex, leftmargin=*]
    \item \textbf{Demographic Parity \cite{KleinbergMR17,DworkHPRZ12}:}
    Demographic Parity measures the difference in the rate at which each class is classified as having an income of higher than 50K.
    \item \textbf{True Positive Rate Balance \cite{Chouldechova17, Simoiu16}:} This measures the overall rate of true positives across both the privileged and non-privileged group.
    \item \textbf{True Negative Rate Balance \cite{Chouldechova17,EqualOpp}:} This measures the overall rate of true negatives across both the privileged and non-privileged group.
    \item \textbf{Conditional Measures \cite{Corbett-DaviesP17}:} We measure the expected value of the above metrics conditioned on the admissible attributes. This more directly captures the notion of justifiable fairness as the admissible variables are directly allowed be used regardless of their discriminatory effects. However, discrimination within groups with identical admissible attributes is prohibited. This measures the discrimination within each group with identical admissible attributes. These measures are abbreviated as their non-conditional counterparts with an additional $C$ prior.
\end{enumerate}

\vspace{-1mm}
 \subsection{Experimental Results}\label{sec:exp-results}

 \common{We first give an overview of all aspects of the results, using \cref{tab:overview}, and then give an in-depth analysis of the quality and fairness results, using \cref{fig:data,fig:fair_compas}. 
 }
 
 \paratitle{Overview of the performance Greedy-\mst}
 \common{
 \cref{tab:overview} gives an overview of the experimental results for Greedy-\sys\ (Algorithm \ref{algo:fair_mst_1}) compared to MST across all datasets with a singular privacy budget of $\epsilon = 1$. Across all datasets, the cost of fairness is relatively low and does not significantly hinder accuracy. 

 In both the Adult and KDD datasets, Greedy-\sys\ incurs no more than a $2\%$ increase in error, on both 1-way and 2-way marginals, due to the additional fairness constraints. Greedy-\sys\ incurs a larger error on the Compas dataset, on the 2-way marginals, incurring an almost $20\%$ increase in error compared to MST. This is a result of the greater inequities that exist in the Compas dataset. In order for \sys\ to satisfy the fairness constraint, the mechanism generates a distribution that has a greater deviation than the fairer datasets. Since each of the mechanisms only differ in how they measure the 2-way marginals this greater deviation is only represented in the 2-way marginals.  This is similarly reflected in the downstream MLP accuracy as Greedy-\sys\ incurs a relatively small penalty in accuracy compared to MST for both the Adult and KDD datasets but incurs a larger penalty for the Compas dataset.
 \par 
 In terms of fairness metrics, Greedy-\sys\ yields better results on all of the fairness measures regardless of the dataset. Since Greedy-\sys\ ensures independence when conditioned on the admissible attributes the conditional fairness metrics are significantly reduced. This also impacts the non-conditional measures as they are also reduced albeit by a smaller amount.
 While all the measures improved by a similar percentage since both the adult and KDD datasets had little disparity originally, the overall change is less significant than the change in the Compas dataset, which has a significantly higher base rate of disparity. We show the significance of these changes in \cref{fig:fair_compas}.\par
 
\common{We have also measured the overall runtime for Greedy-\sys\ compared to MST. Greedy-\sys\ performs slightly better than MST in most cases and the improvement increases with the increase in the size of the dataset (\cref{tab:overview}). This is a result of the greedy optimization where Greedy-\sys\ considers fewer options than MST for neighbors of outcome attributes. We measured the runtime of Exponential-\sys\ as well. Even on small datasets it takes 10 times longer than either MST or Greedy-\sys\ and is prohibitively expensive to run on a large dataset\footnote{Exponential-\sys\ has a significantly higher runtime overhead. \common{As Exponential-\sys\ runs in exponential time we were only able to measure performance on the Adult and Compas datasets.}
\revb{In the low privacy regime} ($\epsilon =0.1$), Exponential-\sys\ can take up to 10 times longer than MST, taking on average 1344 seconds to run. 
In larger datasets, this overhead becomes prohibitively expensive as it scales exponentially in the number of attributes.}}.

}

\paratitle{Data quality}
 \cref{fig:data} shows a detailed breakdown of the quality measures across different privacy budgets and both Greedy and Exponential - \sys.
 Greedy-\sys\ performs similarly to MST  in all cases. In \cref{fig:1way} we can see  that in the worst case when $\epsilon =0.1$ Greedy-\sys\ has only a $1\%$ increase in error. In \cref{fig:2way} we see that this increases to a $7\%$ increase in error across 2-way marginals when compared to MST. 
 While Exponential-\sys\ (Algorithm \ref{algo:fair_mst_2}) performed similarly to MST on 1-way marginals (\cref{fig:1way}), it performed poorly on 2-way marginals (\cref{fig:2way}). At $\epsilon = 0.1$ Exponential-\sys\ sees a $7\%$ increase but when the privacy budget increases to $\epsilon = 10$ that the increase in error is as large as $60\%$.  \reva{Despite the additional optimization in choosing a \mst, Exponential-\sys\ invokes the Gaussian mechanism significantly more often than either MST or Greedy-\sys. This results in less privacy budget for each invocation and thus more noise per measurement leading to more error.}
 
In \cref{fig:Corr}, the patterns are similar to the 2-way marginals. Greedy-\sys\ incurs a small $1\%$ increase in error when the Exponential-\sys\ sees a $56\%$ increase in error at $\epsilon =10$. 
\common{\cref{fig:MLPACC} shows the downstream classification accuracy of MLP models.} \common{The results on the Random Forrest models and linear regression models are largely similar and have been omitted from the figure.} 
 The difference in classification accuracy between MST and Greedy-\sys\ never exceeds $1\%$ even across all classifiers and privacy budgets.
Exponential-\sys\ performs similarly in most cases but performs slightly worse than the baseline in the low privacy setting. When $\epsilon = 10$, the difference between the accuracy of the MLP model trained on MST and Exponential-\sys\ data is $4\%$. 

 \paratitle{Fairness metrics}
 \cref{fig:fair_compas} showcases the fairness measures in more detail for the Compas dataset.
 Every entry in the matrix shows the median results over 10 runs for a specific approach measured by a specific associational fairness metric, where lower numbers (denoted by lighter colors) indicate better fairness.
 Overall, our methods decrease observed unfairness regardless of the downstream classifier used and regardless of the privacy budget. The tradeoff comes in the quality of data. While the downstream classifiers remain fair regardless of the privacy budget the overall accuracy and quality of the data decreases with increased privacy. 
 MST shows significant disparities in both conditional and unconditional measures ($>0.15$). Generating data using either of the fair methods results in a significantly fairer downstream classifier.
 On average, the unconditional fairness metrics of Greedy-\sys\ are $45 \%$ of those seen with MST and the conditional fairness metrics are $20\%$ of those seen with MST. 
 This is expected, as our solutions ensures that there will be no correlation between sensitive attributes and outcome attributes when conditioned on the admissible attributes. This results in a reduction in the unconditional associational metrics while the conditioned metrics are more significantly reduced.
 Of the three classifiers tested, the linear regression model benefited the most from the fair synthetic data.
 On average, when using data from the fair mechanisms the linear regression model saw a decrease in unconditional fairness metrics to $35\%$ of that observed when using the data from MST. The conditional fairness metrics dropped to $14\%$ of what was observed when using data from MST.

\section{Related Work}
We give a detailed review of related works. 
{\em To the best of our knowledge, this is the first work that proposes an approach for generating fair synthetic data while satisfying DP.}

\paratitle{Differentially private data generation}
By releasing synthetic data, one can release a large dataset that can be used for an unbounded amount of computation, all while preserving privacy. There are many methods for generating differentially private synthetic data \cite{ChenOF20,LiuV0UW21,Li21,LiXZJ14,AydoreBKKM0S21,GeMHI21,JordonYS19a,MWEM,LH21,SnokeS18,Torkzadehmahani19,Rosenblatt20,xie18,McKennaSM19,RyanMnist} including methods that use low level marginals to approximate the data \cite{RyanMnist,PrivBays,McKennaSM19}, and GAN based methods \cite{xie18,Rosenblatt20,JordonYS19a} among others. 
According to past work \cite{SynthBenchmark} the marginals-based approaches such as MST \cite{RyanMnist}, PrivBayes \cite{PrivBays} and MWEM-PGM \cite{McKennaSM19} tend to perform best in the private setting. 
Kamino \cite{GeMHI21} also provides DP synthetic data generation with constraints, but focuses on denial constraints \cite{ChomickiM05}, rather than causality-based constraints. 

\paratitle{Fair data generation}
Many works have defined fairness in different ways. Our approach relies on justifiable fairness \cite{InterFair} which is a causal-based notion of fairness.
Other works consider associational fairness measures based on some statistical measures relative to the protected attributes. 
Additionally, conditional associational fairness measures \cite{Corbett-DaviesP17} have been proposed. These measures consider a set of attributes to be conditioned on.
Likewise, there have been approaches for generating non-private fair synthetic data \cite{FairGAN,BreugelKBS21} GAN based approaches such as FairGAN \cite{FairGAN} have been used to generate data that satisfies these associational measures.
While GAN-based approaches work well in the non-private setting, they have been shown to perform poorly in the private setting \cite{SynthBenchmark}.

\section{Extensions of \sys}
\common{Here we detail several possible extensions. These include expanding the techniques used in \sys\ to other private synthetic data mechanisms such as PrivBayes \cite{PrivBays} as well as extending the techniques shown here to numerical and continuous data.}

\paratitle{Other private data generation mechanisms}\label{sec:extensions}
While \sys \ relied on MST~\cite{RyanMnist}, all of our propositions in \cref{sec:model} apply to any Bayes network for synthetic data generation which samples from a generated graphical model, both private and non-private. This applies to private mechanisms beyond MST. For instance, both PrivBayes~\cite{PrivBays} and MWEM-PGM \cite{McKennaSM19} generate a graphical model and sample from it directly. 
To adapt these mechanisms, one only needs to change the selection step, where the graphical model is generated. The selection step must be changed to either \common{consider only admissible attributes as neighbors of outcome attributes} (Algorithm \ref{algo:fair_mst_1}) or find the optimal fair \mst\ (Algorithm \ref{algo:fair_mst_2}).

\paratitle{Additional data types}
\label{sec:categorical}
\revc{MST \cite{RyanMnist} along with other marginal approaches takes discrete data as input.  These techniques can handle numeric attributes by first using a discretization step to map the numeric domain into a discrete domain. This can be done using using a data independent method or a private subroutine such as PrivTree \cite{privtree}. Previous work \cite{SynthBenchmark} has shown that a data dependent discretization can lead to a large increase in performance overall, particularly in large datasets. In these cases, there is a tradeoff between the expressiveness and performance of the dataset. A fine-grained discretization is more expressive but leads to higher noise while coarse-grained discretization loses some information but can result in a lower overall error.
In \cref{sec:experiments}, we evaluate \sys\ on several datasets containing mostly categorical data and discretize numerical attributes with single-value buckets. In future work, it would be intriguing to assess the impact of different discretization approaches on the quality and performance of our system. 
}

\paratitle{Additional fairness definitions}
\common{
Justifiable fairness is only one of many different fairness definitions, both causal and non-causal \cite{KleinbergMR17,DworkHPRZ12,Chouldechova17, Simoiu16,EqualOpp,Corbett-DaviesP17}. While we show in \cref{sec:experiments} that satisfying justifiable fairness also improves additional fairness metrics, there are no guarantees that they are satisfied. As such we leave it to future work to generate private synthetic data that also satisfies additional fairness definitions. Likewise, incorporating integrity constraints in \sys\ is an important future work that will allow the generated data to have other desired properties. While past work \cite{GeMHI21} has incorporated such constraints into private synthetic data, those solutions do not consider 
fairness.}

\section{Conclusion}
We have presented \sys, a system for generating justifiably fair synthetic data under differential privacy. Our approach relies on a state-of-the-art DP data generation system, as well as the definition of justifiable fairness. 
We have formally defined the problem of generating fair synthetic data that preserves DP and showed that it is NP-hard. 
Bearing this in mind, we devised two algorithms that guarantee fairness and DP, and have further extended our greedy approach to other DP synthetic data generation systems. 
Additionally, we have formulated a requirement for the sampling approach to ensure that the generated data comply with the fairness desideratum. 
We have experimentally shown that our approach provides significant improvements in the fairness of the data while incurring low overhead in terms of faithfulness to the private data.

\begin{acks}

This work was supported by the NSF awards IIS-1703431, IIS-1552538, IIS-2008107, and NSF SATC-2016393.
\end{acks}

\clearpage
\balance
\bibliographystyle{ACM-Reference-Format}
\bibliography{bibtex}

\clearpage
\appendix
\section{Proofs}\label{sec:proofs}

\subsection{Proofs for Section \ref{sec:model}}

\begin{proof}[Proof of \cref{thrm:bayes_path}]
The proof is largely the same as that in past work \cite{InterFair}.
Note that for any choice of $K \supseteq A$, if all the directed paths go through at least one attribute in $K$ then intervening on $K$, that is $do(K=k)$ then at least one of the edges in the path will always be severed. This will separate all nodes in $O$ from those in $P$ and thus any further intervention on $P$ cannot influence the distribution of $O$.
\end{proof}

\begin{proof}[Proof of \cref{prop:synth-fair}]
Here, we show that given a set $K \supseteq A$ that the probability of the outcome is dependent only on the intervention on $K$ and is the same regardless of the value of $p$.
We rely on a fact from \cite{pearl95} namely the following equivalence on the do operator. 
\begin{equation}\label{eq:path}
\begin{split}
     P(O=o |do(K=k), do(P=p)) = P(O=o |do(K=k)) \\
    \text{if} (O \perp\!\!\!\perp P |K)_{\overline{PK}} 
\end{split}
\end{equation}

Where $\overline{PK}$ denotes the distribution derived by removing all edges pointing to attributes in $P$ and $K$. \par
This will allow us to show that the outcome is independent of {\em any intervention on $P$} once we have already intervened on $K$,i.e., applying the $do$ operator on $\prot$ does not change the probability. 
In particular, we will be able to show that $P[O=o|do(P_i=0), do(K=k)]=  P[O=o|do(P_i=1), do(K=k)] = P[O=o|do(K=k)]$, which is the definition of K-fairness.

We also highly rely on the concept of d-separation a notion that can be directly observed on the graph. We say two attributes $X$ and $Y$ are d-connected (and thus not d-separated) conditioned on a set of attributes $Z$ if any of the following hold. 
\begin{itemize}
    \item There exists an undirected path from $X$ to $Y$ that is not blocked by $Z$
    \item There exists an undirected path from $X$ to $Y$ that is not blocked by a collider $a \notin Z$
\end{itemize}
Where a collider is an attribute along the undirected path where both edges in the path are directed towards that attribute.

Let $K$ be any superset of $A$ which shares no elements with $P$. Note that in any directed \mst there is exactly one undirected path between any attribute in $P$ and any attribute in $O$. Likewise this path is always blocked by at least one attribute in $K$.

\begin{figure}[t]
\begin{center}
\begin{tikzpicture}[scale=0.6]
\begin{scope}[every node/.style={shape=rectangle, rounded corners,thick,draw,, top color=white, bottom color=blue!25}, 
other/.style={shape=rectangle, rounded corners,thick,draw,, top color=white, bottom color=red!25},
outcome/.style={shape=rectangle, rounded corners,thick,draw,, top color=white, bottom color=green!25},
none/.style={shape=rectangle, rounded corners,thick,draw,, top color=white, bottom color=light-gray}]
    \node[other] (pi) at (0,0) {\footnotesize $\Pi$};
    \node[none] (X) at (2.5,0) {\footnotesize $X$};
    \node (A) at (5,0) {\footnotesize $\alpha$};
    \node[none] (Y) at (7.5,0) {\footnotesize $Y$};
    \node[outcome] (O) at (10,0) {\footnotesize $\Omega$};
    
    \node[other] (pi2) at (0,-2) {\footnotesize $\Pi$};
    \node[none] (X2) at (2.5,-2) {\footnotesize $X$};
    \node (A2) at (5,-2) {\footnotesize $\alpha$};
    \node[none] (Y2) at (7.5,-2) {\footnotesize $Y$};
    \node[outcome] (O2) at (10,-2) {\footnotesize $\Omega$};
    
    \node[other] (pi3) at (0,-4) {\footnotesize $\Pi$};
    \node[none] (X3) at (2.5,-4) {\footnotesize $X$};
    \node (A3) at (5,-4) {\footnotesize $\alpha$};
    \node[none] (Y3) at (7.5,-4) {\footnotesize $Y$};
    \node[outcome] (O3) at (10,-4) {\footnotesize $\Omega$};
\end{scope}

\begin{scope}[>={Stealth},
            standard/.style={draw=black, thick}, 
             unfair/.style={draw=red, thick}]
    \path [-,standard,<->] (pi) edge node {} (X);
    \path [-,standard,->] (X) edge node {} (A);
    \path [-,standard,->] (A) edge node {} (Y);
    \path [-,standard,<->] (Y) edge node {} (O);
    
    \path [-,standard,<->] (pi2) edge node {} (X2);
    \path [-,standard,->] (X2) edge node {} (A2);
    \path [-,standard,<-] (A2) edge node {} (Y2);
    \path [-,standard,<->] (Y2) edge node {} (O2);
    
    \path [-,standard,<->] (pi3) edge node {} (X3);
    \path [-,standard,<-] (X3) edge node {} (A3);
    \path [-,standard,->] (A3) edge node {} (Y3);
    \path [-,standard,<->] (Y3) edge node {} (O3);
\end{scope}
\end{tikzpicture}
\end{center}
    \caption{Possible paths for the proof of \cref{prop:synth-fair}.}
    \label{fig:paths}
\end{figure}
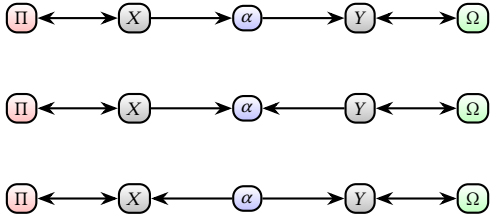

Therefore the path from the attribute in $P$ to an attribute in $O$ must have one of the structures shown in \cref{fig:paths} where $X$ and $Y$ denote arbitrary sets of nodes along the path with arbitrary directed edges between them. Double edges denote an edge that can be in any arbitrary direction. $\Pi$ denotes an attribute in $P$, $\Omega$ denotes an attribute in $O$ and $\alpha$ denotes an attribute in $A$. Note that the path must either pass through $\alpha$ (the first case), have edges that go to $\alpha$ (the second case) or have edges that originate from $\alpha$ (the third case). \par 
In the graph  $\overline{PK}$ where all edges going into attributes in $P$ and $K$ are removed each of these cases introduces a d-separation between $\Omega$ and $\Pi$ when conditioned on $K$. In the first case, the edge going into $\alpha$ is removed and as such the path is severed. In the second case, both edges going to $\alpha$ are removed and the path is severed. In the third case, no edges are severed but conditioning on $\alpha$ d-separates $\Pi$ and $\Omega$ since it blocks the path and is not a collider. \par 

Since the dataset is Markov compatible with the graph then the d-separation of $\Omega$ and $\Pi$ imply that the two are independent when conditioned on $\alpha$. Since they are conditionally independent we can use \cref{eq:path} to show that $P(O=o|do(K=k), do(P=p))$ is the same regardless of the intervention on $P$. Therefore, the distribution satisfies K-fairness. Likewise since $K$ could be an arbitrary superset of $A$ then the distribution also satisfies justifiable fairness.
\end{proof}

\begin{proof}[Proof of \cref{prop:nphard}]
We show that, given an instance of 3-SAT, we can define an equivalent instance of the decision version of \prob, where any maximal spanning tree will correspond to a satisfying 3-SAT assignment. \par 

\paratitle{Construction}
First we describe the construction of a graph $\gG$ from a 3-SAT instance $\varphi$ with $m$ clauses and $n$ literals.
We create an assignment gadget (\cref{fig:assign}) for each literal in $\varphi$. This contains a protected attribute $\Pi_i$ and edges from the protected attribute to two additional attributes representing the literal and its negation ($x_i$ and $\neg x_i$ respectively). For each clause in $\varphi$ we create a 3-way OR gadget (\cref{fig:3_OR}). The $\alpha$ nodes in the OR gadget are admissible attributes and the $\Omega$ nodes are outcome attributes. Each of the inputs for the 3-way or gate ($x_1,x_2,x_3$) are then connected to the corresponding literal in the assignment gadget using a weight 3 edge. 
We now prove that there is a solution to the decision version of \prob\ (\cref{def:dec_prob}) with $k = 22m + 2n$ 

To build the intuition behind the 3-way OR gadget we first describe the 2-way OR gadget (\cref{fig:2_OR}) and build upon that. This gadget takes two inputs $x_1$ and $x_2$. These may be either connected to an assignment gadget or the output of another gadget. Note that in order to make the maximum spanning tree across this substructure both edges with weight 2 must be taken and at least one edge with weight 1 must be taken resulting in a total weight of $5$. We recall that we consider a literal to be set to $False$ if there exists an unblocked path from a protected attribute to it. We now consider the output of the two way OR gadget $O$. This node will be considered set to $False$ if there is an unblocked path between a protected attribute and $O$. For either of the possible maximum spanning trees on the 2-way OR gadget only one of the inputs can have a path to $O$ that is blocked by $\alpha$. 
Therefore, for either of the maximum spanning trees, if both inputs are set to $False$ only one of them may be blocked by $\alpha$ and the other will have a direct edge to the output resulting in an output of $False$. As such any maximum spanning tree on the 2-way OR gadget evaluates to $True$ if at least one of its inputs evaluates to $True$ resulting in the same functionality as an OR gate. Similar to traditional OR gates in order to construct the 3-way OR gadget we simply apply two 2-way OR gadgets together with the only exception being that the output of the second 2-way or gadget is an outcome attribute. This ensures that if a tree over the 3 way or gadget  has the maximum weight (13) and satisfies \cref{thrm:bayes_path} then it corresponds to a satisfying assignment for the corresponding clause.

iff $\varphi$ has a satisfying assignment.

\paratitle{Proof}
($\Rightarrow$) Assume that we have a fair \mst\ $T$ of the constructed graph $G$. 
If $T$ has the maximum weight of $22m + 2n$, where $m$ is the number of clauses and $n$ is the number of literals, and then the assignment to each literal can be read by looking at each assignment gadget. In $T$ only one of the edges from each assignment gadget will be kept due to the tree condition. Whichever node has an edge to the protected attribute $\Pi_i$ will be set to $False$ and the other will be set to $True$. This assignment is consistent as the literal in the assignment gadget in \cref{fig:assign} will have an edge to every 3-way OR gadgets that corresponds to clauses which have that literal. This ensures that all 3-way OR clauses which have $x_1$ will have an edge to $x_1$ in the assignment gadget and those that have $\neg x_1$ have an edge to $\neg x_1$ in the assignment gadget. Since only one edge can be taken in the assignment gadget then either $x_1$ or $\neg x_1$ can be set to $False$ but not both. \par 
Recall that in order for the minimum spanning tree over the 3-way OR gadget to be maximum and fair at least one literal must be set to $True$. Because of this any global \mst of the maximum weight of  $13m + 5n$ ensues that each clause has at least one literal set to $True$. 

($\Leftarrow$) Here we show that given a satisfying assignment to $\varphi$, $a$, we can construct a unique fair \mst\ on $G$.
In order to show that a minimum spanning tree of $G$ always corresponds to a 3-SAT solution we will rely heavily on the following fact about MSTs: for any spanning tree on graph $G$ to be an MST all the subtrees on smaller connected components must also be a minimum spanning trees for each of those components. This allows us to reason about the smaller parts of the tree (mainly the OR gadget) without considering the rest of the tree.\par 

\begin{figure}
    \begin{center}
\begin{tikzpicture}[scale=0.6]
\begin{scope}[every node/.style={shape=rectangle, rounded corners,thick,draw,, top color=white, bottom color=blue!25}, 
other/.style={shape=rectangle, rounded corners,thick,draw,, top color=white, bottom color=red!25},
none/.style={shape=rectangle, rounded corners,thick,draw,, top color=white, bottom color=light-gray},
outcome/.style={shape=rectangle, rounded corners,thick,draw,, top color=white, bottom color=green!25}]
    \node[none] (X1) at (0,5.5) {\footnotesize $x_1$};
    \node (A) at (2.7,4.2) {\footnotesize $\alpha$};
    \node[none] (X2) at (0,3) {\footnotesize $x_2$};
    \node[none] (O) at (5,4.2) {\footnotesize $\outcome$};
\end{scope}

\begin{scope}[>={Stealth},
            standard/.style={draw=black, thick}, 
              unfair/.style={draw=red, thick}]
    \path [-,standard,-] (X1) edge node[above=2pt] {$2$} (A);
    \path [-,standard,-] (X2) edge node[below=2pt] {$2$} (A);
    \path [-,standard,-,bend left=20] (X1) edge node[above=2pt] {$1$} (O);
    \path [-,standard,-,bend right=20] (X2) edge node[below=2pt] {$1$} (O);
\end{scope}
\end{tikzpicture}
\end{center}
    \caption{2-way OR gadget.}
    \label{fig:2_OR}
\end{figure}
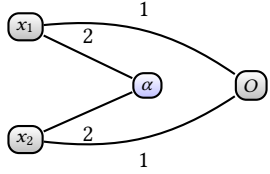

Now given a satisfying assignment to $\varphi$, $a$, we can construct the corresponding fair \mst. First, select the edge that correctly assigns  the attributes according to the assignment gadget. That is if the assignment sets $x_1$ to $True$ select $(\Pi\neg,x_1)$  in \cref{fig:assign} and   $(\Pi, x_1)$ otherwise. Then for each 3-way OR gadget (\cref{fig:3_OR}) take all four weight 2 edges. 

\begin{figure}
    \begin{center}
\begin{tikzpicture}[scale=0.6]
\begin{scope}[every node/.style={shape=rectangle, rounded corners,thick,draw,, top color=white, bottom color=blue!25}, 
other/.style={shape=rectangle, rounded corners,thick,draw,, top color=white, bottom color=red!25},
outcome/.style={shape=rectangle, rounded corners,thick,draw,, top color=white, bottom color=green!25},
none/.style={shape=rectangle, rounded corners,thick,draw,, top color=white, bottom color=light-gray}]
    \node[none] (X1) at (0,4.9) {\footnotesize $x_1$};
    \node (A) at (2.7,3.9) {\footnotesize $\alpha$};
    \node[none] (X2) at (0,3.1) {\footnotesize $x_2$};
    \node[none] (O) at (5,3.9) {\footnotesize $\outcome$};
    
    \node[none] (X') at (7,3.9) {\footnotesize $x'$};
    \node (A2) at (9.7,2.9) {\footnotesize $\alpha$};
    \node[none] (X3) at (7,2) {\footnotesize $x_3$};
    \node[outcome] (O') at (12,2.9) {\footnotesize $\Omega$};
\end{scope}

\begin{scope}[>={Stealth},
            standard/.style={draw=black, thick}, 
              unfair/.style={draw=red, thick}]
    \path [-,standard,-,draw=blue] (X1) edge node[above=0pt] {$2$} (A);
    \path [-,standard,-,draw=blue] (X2) edge node[below=0pt] {$2$} (A);
    \path [-,standard,-,bend left=20,draw=blue] (X1) edge node[above=2pt] {$1$} (O);
    \path [-,standard,-,bend right=20] (X2) edge node[below=2pt] {$1$} (O);
    
    \path [-,standard,-,draw=blue] (X') edge node[above=0pt] {$2$} (A2);
    \path [-,standard,-,draw=blue] (X3) edge node[below=0pt] {$2$} (A2);
    \path [-,standard,-,bend left=20,draw=blue] (X') edge node[above=2pt] {$1$} (O');
    \path [-,standard,-,bend right=20] (X3) edge node[below=2pt] {$1$} (O');
    
    \path [-,standard,-,draw=blue] (O) edge node[above=2pt] {$3$} (X');
\end{scope}

\end{tikzpicture}
\end{center}
    \caption{Example 3-Way OR gadget. The satisfying edges are highlighted in blue }
    \label{fig:SAT}
\end{figure}
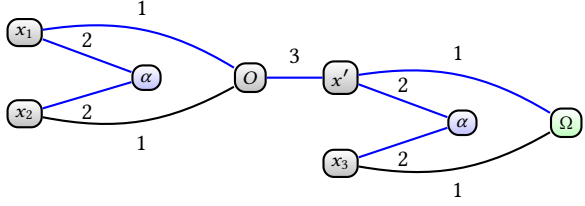

In the case of the 3-way OR gadget all edges will refer to the notation of \cref{fig:3_OR}. Since $a$ is a satisfying assignment for $\varphi$, at least one of the 3 literals $x_i$ will evaluate to $True$ . If this literal is connected to the first 2-way OR gadget , select its 1 weight edge to be connected to the output $(x_i,O)$ and the 1 weight edge from the output to the outcome node $(x', \Omega)$. If the literal is connected to the second 2-way OR gadget either weight 1 edge $(x_1, O)$ or $(x_2,O)$ from the first 2 weight or gadget can be selected and the weight 1 edge from the literal to the outcome $(x_3, \Omega)$ will be selected in the second 2-way OR gadget. We give an example of such a 3-WAY OR gadget in \cref{fig:SAT} where the edges to be taken are highlighted in blue. This corresponds to the assignment $(x_1 = True, x_2 = False, x_3 = False) $. Taking these edges both maximizes the weight as it takes all the 2 weight edges and only one 1 weight edge from either 2-way or gadget. This also ensures that the resulting \mst\ is fair since each outcome in the 3-way OR gadget has paths all blocked by admissible attributes. In each case for any maximal spanning tree at most one path from the literals to the outcome may be blocked since only one of the weigh 1 edges can be taken. 

\end{proof}

\begin{proof}[Proof of \cref{lemma:database_existance}]
Here we first show that we can create a set of random variables $A,B,C,D$ which results in the graph where $B, C,D$ all have arbitrary mutual information with $A$ and there is no mutual information between any other pair of variables. From there we demonstrate that we can extend this construction to an arbitrary number of variables  each with their arbitrary weight. We argue that this construction is sufficient to be able to construct a graph with the structure of the 3-way OR gadget.  \par 
Let $B,C$ and $D$ be discrete uniform random variables with domain size $n$ for some $n \in \mathbb{N}$. Let $A$ be the discrete random variable with the following probability distribution. 
\[
\begin{cases}
A = B&\text{With Probability $x$}\\
A = C&\text{With Probability $y$}\\
A = D&\text{With Probability $z$}\\
A = 0&\text{With Probability $\lambda$}
\end{cases}
\]
Where $x+y+z+\lambda = 1$.
The entropy of $A$ from is as follows. 
\begin{equation}
    \begin{split}
        H(A) = & \\
        & -(x \log(x/n) + y \log(y/n) + z \log(z/n) + (\lambda)log(\lambda))
    \end{split}
\end{equation}
Likewise the conditional entropy is as follows.
\begin{equation}
    \begin{split}
        H(A|B) = & \\
        & -(x \log(x) + y \log(y/n) + z \log(z/n) + (\lambda)log(\lambda))
    \end{split}
\end{equation}
As such the mutual information is as follows. 
\begin{equation}
    \begin{split}
        I(A;B) = H(A)-H(A|B)= & \\
        & x \log(x) - x\log(x/n) = \\
        & x(\log(x) - \log(x/n)) = \\
        & x \log(n)
    \end{split}
\end{equation}

The same can be done for the other random variables $B$ and $C$. It is clear that $B,C$ and $D$ have mutual information $0$ since they are all uniform random variables. 
\par  The only constraint is that $x + y + z + \lambda =1$. Given a sufficiently large $n$ we can set $x\log(n)$ to be any arbitrary value and the same goes for the other two attributes. For example if for example we would like to make the relationship between $O, x', \alpha$ and $\Omega$ shown in \cref{fig:3_OR} we would do the following construction. \par 
Let $A$ take the place of $x'$, $B$ take the place of $O$, C take the place of $\alpha$ and $D$ take the place of $\Omega$. Let $n = 2^{12} =4096,x = \frac{1}{4}, y = \frac{1}{6}, z = \frac{1}{12} $ and $\lambda = \frac{1}{2}$. First these values of $x, y, z $ and $\lambda$ satisfy the constraint as $\frac{1}{4} + \frac{1}{6} +\frac{1}{12} +\frac{1}{2} = 1$. Likewise these probabilities give us the desired mutual probabilities as $x \log (n) = \frac{1}{4} \log (4096) = 3$. Likewise $y \log (n) = 2$ and $z \log (n) = 1$ which is what we need for $A = x'$, $B = \alpha$, $C = \Omega$, and $D = O$ in \cref{fig:3_OR}. \par 
We can extend this to any arbitrary number of edges as well as any value of mutual information. This only requires a larger value of $n$. If we would like $\zeta$ edges with the maximum mutual information $I$ $n$ must be at least $2^{\zeta I}$

\end{proof}

\subsection{Proofs for Section \ref{sec:algorithms}}

\begin{proof} [Proof of \cref{prop:exp_rdp}(1)]
Using the Gaussian Mechanism (line \ref{line:exp_2_way}) with scale $\sigma$ satisfies $(\alpha ,\frac{1}{2\sigma ^2})$-RDP. If we set $\sigma = \sqrt{\frac{r}{2\rho}}$ this becomes $(\alpha, \alpha \frac{\rho}{r})$-RDP. 
 This is invoked $r$ times and is the only line that directly accesses private data thus by \cref{thrm:RDPcomposoition} it thus satisfies $(\alpha, \alpha \rho)$-RDP.
\end{proof}

\begin{proof} [Proof of \cref{prop:exp_rdp}(2)]
When Algorithm \ref{algo:fair_mst_2} creates incomplete spanning tree to add to the priority queue it rejects any partial spanning tree that violate \cref{thrm:bayes_path}. Therefore the resulting output satisfies \cref{thrm:bayes_path}.
\end{proof}

\begin{proof}[Proof of \cref{prop:exp_time}]
Let $G$ be a graph over the input database $D$ where the nodes represent attributes and edges represent mutual information between attributes. Consider the case where each edge has uniform weight (by \cref{lemma:database_existance} such a database exists). In this case, the weight of any partial spanning tree  is merely the number of edges in the partial tree. As such, Algorithm \ref{algo:fair_mst_2} will create every possible spanning trees before selecting any of them . There are $|\gA|^{|\gA| -2}$ possible spanning trees resulting in the worst case time complexity.
\end{proof}

\begin{proof}[Proof of \cref{prop:greedy_rdp}(1)]
Line 8 satisfies $(\alpha, \alpha \frac{1}{8}\epsilon^2)$-RDP. This is the only line that accesses the private data. If we set $\epsilon = \sqrt{\frac{8\rho}{r-1}}$ this becomes $(\alpha, \alpha \frac{\rho}{r-1})$-RDP. This is invoked $r-1$ times and by \cref{thrm:RDPcomposoition} it thus satisfies $(\alpha, \alpha \rho)$-RDP
\end{proof}

\begin{proof}[Proof of \cref{prop:greedy_rdp} (2)]
In line \ref{line:row_remove}, Algorithm \ref{algo:fair_mst_1} removes any edges from nodes in $\outcome$ that do not go to nodes in $\outcome \cup \admissible$. This results in a \mst\ where all of the neighbors of nodes in $\outcome$ are in $\outcome \cup \admissible$. Any path from attributes in $\protect$ to attributes in $\outcome$ must pass through a neighbor of the attribute in $\outcome$ and therefore must pass through an attribute in $\admissible \cup \outcome $. Therefore by \cref{prop:fair_parents} the output of Algorithm \ref{algo:fair_mst_1} is a fair \mst.
\end{proof}

\begin{proof} [Proof of \cref{prop:fair_parents}]
Assume that all the neighbors of nodes in $\outcome$ are in $\outcome \cup \admissible$. Any path from a node in $\prot$ to a node in $ o \in \outcome$ must pass through at least one of $o$'s neighbors. If that neighbor is in $\admissible$ then the path is blocked by an admissible attribute therefore satisfying \cref{thrm:bayes_path}. If that neighbor was in $\outcome$ then the path must also go through one of its neighbors which must be in $\outcome \cup \admissible$. Since only nodes in $\admissible$ are allowed to have edges to nodes outside of $\outcome \cup \admissible$ then the path must go through at least one of these nodes therefore blocking the path and satisfying \cref{thrm:bayes_path}.   \par 
To prove the converse we use a proof by contradiction. Assume $T$ is a fair \mst, the problem setting is saturated,  and that there is an edge from a node $o \in \outcome$ to a node $x \notin \outcome \cup \admissible$. By the definition of the saturated case since $x$ is not in $\outcome \cup \admissible$ then it must be in $\prot$. Since $x$ is in $\prot$ and is neighbor of $o$ then there exists a direction of edges ($x$->$o$) such that $o$ is directly dependent on $x$ resulting in a violation of \cref{def:fair-mst}.

\end{proof}

\begin{proof}[Proof of \cref{prop:greedy_time}]
Algorithm \ref{algo:fair_mst_1}  greedily selects $|\gA|-1$ edges one at a time using the exponential mechanism (lines \ref{line:start_for}-\ref{line:end_for}). This process is done exactly $|\gA|-1$ times resulting in a time complexity of $|\gA|-1$.
\end{proof}


\end{document}